**FRIEDRICH-ALEXANDER-UNIVERSITÄT ERLANGEN-NÜRNBERG**

Lehrstuhl für VWL, insbes. Arbeitsmarkt- und Regionalpolitik
Professor Dr. Claus Schnabel

# The Myth of Restrictive Working-Time Agreements. Firms' Working-Time Adjustment after Exit from Collective Bargaining

VINZENZ PYKA AND ANDRE RIEDER

# The Myth of Restrictive Working-Time Agreements. Firms' Working-Time Adjustment after Exit from Collective Bargaining*

Vinzenz Pyka† and André Rieder‡

**Abstract:** Employers in Germany argue that collective agreements restrict their flexibility concerning working hours. Using data from the IAB Establishment Panel for the years 2008 to 2024, we investigate working-time adjustments after exit from collective bargaining. We find that exit is associated with only a modest increase in weekly working hours of around 0.3 to 0.6 per cent using fixed-effects and event-study models, corresponding to eight to thirteen additional minutes per week. This adjustment occurs shortly after leaving collective bargaining coverage. We further find no evidence that other margins of working-time adjustment, such as Saturday or Sunday work, are used more extensively after exit. In contrast to previous survey evidence, our findings call into question that working-hours constraints are one of the main drivers of exit from collective bargaining coverage.

**Zusammenfassung:** Arbeitgeber in Deutschland argumentieren häufig, dass tarifvertragliche Regelungen ihre Flexibilität bei der Festlegung der Arbeitszeitdauer einschränken. Auf Grundlage des IAB-Betriebspanels 2008–2024 untersuchen wir, wie Betriebe ihre Arbeitszeit nach dem Austritt aus der Tarifbindung anpassen. Fixed-Effects- und Event-Study-Schätzungen zeigen lediglich einen moderaten Anstieg der vereinbarten Wochenarbeitszeit um 0,3 bis 0,6 Prozent. Dies entspricht einer Verlängerung der wöchentlichen Arbeitszeit um etwa acht bis dreizehn Minuten. Bemerkenswert ist, dass die Anpassung trotz der Nachwirkung tarifvertraglicher Regelungen überwiegend kurz nach dem Austritt erfolgt. Für andere Anpassungen, etwa Samstags- oder Sonntagsarbeit, finden wir keine Hinweise. Damit stellen unsere Ergebnisse bisherige Befunde aus Unternehmensbefragungen infrage, wonach tarifvertragliche Arbeitszeitregelungen zu den Hauptgründen für einen Austritt aus der Tarifbindung zählen.

---

* The authors thank Claus Schnabel, Stephanie Prümer, Matthias Collischon, Mario Bossler, and Thomas Vollmuth for valuable feedback, and Pascal Beisenwenger for excellent research assistance.

† Friedrich-Alexander-Universität Erlangen-Nürnberg, Nürnberg, Germany; correspondence to: Vinzenz Pyka, University of Erlangen-Nürnberg, School of Business, Economics and Society, Lange Gasse 20, 90403 Nürnberg, Germany. E-Mail: vinzenz.pyka@fau.de

‡ Friedrich-Alexander-Universität Erlangen-Nürnberg, Nürnberg, Germany;

## 1 Introduction

Across[1] OECD countries, working time is an important subject of collective bargaining. However, the scope for firm-level adjustments varies considerably depending on the bargaining system in place. A limited ability to adjust collective bargaining regulations, such as those relating to working time, shifts and leave, may restrict firms' flexibility and reduce productivity (OECD, 2019, 2021). Consequently, employers often perceive working time regulations in collective bargaining agreements as restrictive. Survey evidence from Germany indicates that working-time provisions are among the main reasons why firms leave collective bargaining coverage (Langer, 1994; Lehmann, 2002; Lesch et al., 2018).

During the 2018 bargaining round in Germany's metal industry, employers' associations raised concerns that the agreement negotiated between Germany's largest trade union, IG Metall, and the corresponding employer association Südwestmetall could exacerbate existing shortages of skilled labour. The settlement allowed workers to reduce their working time per week from 35 to 28 hours while preserving the right to return to full-time work. Stefan Wolf, Südwestmetall's chief negotiator, described it as "a burden which will be hard to bear for many firms". Trade unions, by contrast, have historically campaigned for higher wages and shorter working hours and reject this argument. During the 2018 bargaining round, IG Metall chairman Jörg Hofmann stated "We want shorter hours to be possible for everyone, including those at the bottom of the income scale" (Chazan, 2018; Wissenbach, 2018).

This German case highlights a wider, ongoing debate in industrial relations, raising two key questions. First, do collective bargaining agreements restrict employers' flexibility with regard to working hours? Second, are working-time regulations an important driver of establishments' decisions to leave collective bargaining coverage?

In our study, we shed light on the relationship between establishments' exit from collective bargaining coverage and subsequent adjustments in working time. Given that working time is one of the main margins for adjusting labour input, it is surprising that empirical research on establishments' exit from collective bargaining coverage has focused predominantly on wages (e.g. Addison et al., 2016; Bossler, 2019; Gürtzgen, 2016; Hirsch & Mueller, 2020).

[1] Artificial intelligence, like ChatGPT, was utilised to enhance the text by refining its grammar and sentence structure.

Our analysis is based on the presumption that if collective bargaining agreements substantially restrict establishments' flexibility, leaving coverage should be associated with an increase in working hours or greater use of other instruments for adjusting working time. To examine this relationship, we use data from the IAB Establishment Panel for the period 2008 to 2024 and apply two-way fixed-effects and event-study estimators. We find an increase of approximately 0.3 to 0.6 per cent in agreed weekly working hours for full-time employees, corresponding to around eight to thirteen additional minutes per week, which amounts to roughly one additional working day per year. At the same time, we find only small and statistically insignificant changes in the use of Saturday and Sunday work as instruments for adjusting working time. These findings suggest that collective bargaining constrains contractual working hours to some extent, but that leaving collective bargaining coverage does not result in a substantial reorganisation of regular weekly working time.

We contribute to the literature in two respects. First, we extend research on employers' withdrawal from collective bargaining coverage, which has focused predominantly on wages while paying comparatively little attention to adjustments in contractual working hours. Second, we contribute to the literature on working time by providing new evidence on how contractual working hours change when establishments leave collective bargaining coverage. This relationship has received little attention in recent research and has mainly been examined in older studies (Kölling & Lehmann, 2002). More broadly, our findings challenge the view that collective bargaining either simply restricts managerial flexibility or leaves firms entirely unconstrained. Instead, they indicate that collective agreements impose limited constraints on the duration of contractual working time.

Our findings are particularly relevant in the context of Germany's weak economic growth, rising vacancy-to-job-seeker ratio, declining collective bargaining coverage, and decreasing average annual hours worked per worker (Bossler & Popp, 2026; Hohendanner & Kohaut, 2025; Kölling, 2022; OECD, 2025). These developments have contributed to a broader political debate about labour supply and working-time flexibility, including proposals to reform Germany's Working Time Act by replacing the daily maximum working time with a weekly limit (Deutscher Bundestag, 2026). While employers' associations have broadly welcomed such proposals (BDA, 2025), trade unions argue that collective bargaining agreements already provide an appropriate framework for regulating working time in Germany and warn that established worker protections could be weakened (DGB, 2025, 2026).

The remainder of the article proceeds as follows. The next section outlines the institutional framework and develops the theoretical expectations. The subsequent sections introduce the data and empirical strategy, present the empirical findings, and discuss their implications for research on collective bargaining and labour-market flexibility.

## 2 Institutional background

The institutional framework governing working time in Germany is defined by the Working Time Act (*Arbeitszeitgesetz*, ArbZG), enacted in 1994. The Act sets binding upper limits on daily working hours, mandates minimum rest periods, and restricts work on Sundays and public holidays (for further details, see Bispinck (2006)). It limits working time to eight hours per working day, which may be extended to up to ten hours provided that average working time does not exceed eight hours per working day within six calendar months or 24 weeks. Since Saturday counts as a working day, this corresponds to a statutory maximum of 48 hours per week. In practice, however, the five-day week with about 40 hours per week is the standard in the German labour market, reflecting, among other factors, the historical influence of trade unions.[2]

While statutory law defines minimum standards, the organisation of working time in practice is shaped at multiple levels. In principle, working time, like wages, forms part of the individual employment contract and can therefore be negotiated at the firm or individual level. In the German institutional setting, however, working-time arrangements are commonly regulated collectively through establishment-level works agreements (*Betriebsvereinbarungen*) or collective bargaining agreements between employee representatives, such as trade unions and employers at the sectoral or establishment level (Schnabel, 2025).

Collective agreements allow for sector- and firm-specific adaptations within the statutory framework and often go beyond legal minimum requirements. As documented by Bispinck (2006, 2025), collectively negotiated standards contribute to substantial heterogeneity in working-time arrangements across sectors. Under certain conditions, collective agreements may be declared generally binding by public authorities. In such cases, their provisions, including those governing wages and working time, also apply to establishments that were previously not covered by the agreement within the relevant sector (Schnabel, 2025).

[2] The German Trade Union Confederation (DGB) began campaigns for the 40-hour working week in the 1950s.

Beyond setting working-time, collective agreements also regulate the distribution and flexibility of working hours. As described by Bispinck (2006), they typically include detailed provisions on overtime, compensation schemes and temporal flexibility, including instruments such as working-time corridors and working-time accounts. These mechanisms allow firms to adjust working hours within predefined limits or to smooth fluctuations over time without immediately triggering overtime compensation. Collective bargaining therefore not only constrains firms but also provides structured forms of internal flexibility.

Despite these flexibility-enhancing features, collectively agreed standards may nevertheless limit firms' ability to respond to changing economic conditions, particularly during downturns or periods of structural change. Firms can respond to such constraints in several ways: they may refrain from entering collective agreements, renegotiate existing agreements or exit collective bargaining coverage altogether. Importantly, exit from collective bargaining does not immediately eliminate the effects of previously applicable agreements. Under the Collective Agreements Act (*Tarifvertragsgesetz*, TVG), the normative provisions of expired collective agreements continue to apply until they are replaced by another (individual) agreement, a principle known as *Nachwirkungspflicht* in Germany. [3]

## 3 Existing literature

Studies use working time in different context such as labour supply, well-being, health, gender inequality as well as working-time preferences as an explanatory variable (e.g. Cygan-Rehm & Wunder, 2018; Erosa et al., 2022; Wunder & Heineck, 2013). However, as an dependent variable considerably less attention has been paid to working-time adjustment in the context of collective bargaining. This leaves an important gap in the literature, particularly given the central role that collective agreements play in regulating working time in coordinated labour markets.

Collective agreements define standard working hours, regulate overtime and compensation, and provide negotiated instruments such as working-time accounts, corridors and sector-specific adjustment rules. Yet the empirical literature has paid comparatively little attention to whether collective agreements actually constrain such adjustments, or whether establishments use regained autonomy after bargaining exit to alter working-time arrangements.

---

[3] The obligation of continued applicability is regulated in § 4 of the Collective Agreements Act (Tarifvertragsgesetz).

In Germany, research on changes in collective bargaining status has focused primarily on wages rather than working time. Existing studies show that exits from collective bargaining are associated with wage reductions and that establishments orienting their wage-setting towards collective agreements pay lower wages than formally covered establishments (Addison et al., 2016; Bossler, 2019; Gürtzgen, 2016; Hirsch & Mueller, 2020). By contrast, evidence on whether establishments also adjust working time after leaving collective bargaining coverage remains scarce.

This topic is relevant because survey and case-study evidence suggests that working-time regulations are an important reason why firms leave employers' associations and thus collective bargaining coverage. In the German metal industry, Langer (1994) shows that reductions in standard working hours, particularly the introduction of the 35-hour week, played a key role in firms' exit decisions. Similar evidence is reported by Lehmann (2002) and Lesch et al. (2018), who show that dissatisfaction with both wage levels and working-time provisions is central to firms' decisions to leave collective bargaining coverage. These findings suggest that employers perceive working-time provisions as relevant constraints. However, they do not show whether establishments actually adjust working time after leaving collective bargaining coverage.

To our knowledge, the only closely related evidence for Germany is provided by Kölling and Lehmann (2002) who examine establishments leaving collective bargaining coverage between 1996 and 1999. They report an increase in weekly working hours of approximately 30 minutes in West Germany and ten minutes in East Germany. However, our study differs decisively from their study in multiple ways: First, we are using more advanced methods of econometric analysis such as stacked event studies who have established in the literature in recent years. This allows us to trace the timing and persistence of post-exit adjustments of working time after collective bargaining exit. Second, we are using from the IAB establishment panel for the period 2008 to 2024. This allows us to examine whether these findings of by Kölling and Lehmann (2002) also hold in a more recent period characterised by substantially lower bargaining coverage and a broader use of negotiated working-time flexibility. Thirdly, we can examine sectoral heterogeneity and analyse other aspects of working time organisation, such as the use of Saturday and Sunday working, and working time corridors. A dimension not taken into account by Kölling and Lehmann (2002).

Our study therefore provides new evidence on whether the association between bargaining exit and working time observed in the late 1990s persists in a substantially transformed collective bargaining system. We also broaden the literature on employer exit from collective bargaining by shifting attention from wages to working-time regulation. In doing so, we assess whether collective agreements primarily constrain establishments' working-time flexibility or instead provide a negotiated framework within which such adjustment is already possible.

## 4 Data

To examine the relationship between establishments' exits from collective bargaining coverage and working time, we draw on the IAB Establishment Panel, a widely used longitudinal dataset for establishment-level research in Germany (e.g. Bossler & Gerner, 2019; Rieder & Schnabel, 2026). Conducted annually since 1993, the panel covers establishments across all sectors of the German economy. Its sampling frame comprises all establishments employing at least one worker subject to social security contributions as of 30 June in a given year. Approximately 15,000 to 16,000 establishments are surveyed each year, with a substantial panel component enabling repeated observation of the same units over time. The sampling design is based on disproportionate stratification by establishment size, industry, and federal state, ensuring sufficient representation across key structural dimensions. Regular sample refreshment compensates for panel attrition and helps maintain representativeness. Data are collected primarily through face-to-face interviews with establishment owners or senior managers by professional interviewers. The panel is highly stable, with a retention rate of around 84 per cent among establishments participating in the previous wave. Further details are provided by Ellguth et al. (2014), Bossler et al. (2018) and Bellmann et al. (2026).

The IAB Establishment Panel is particularly suited to our analysis because it combines detailed information on collective bargaining coverage with establishment-level measures of working time. The data distinguish between establishments not covered by collective agreements, those covered by sectoral agreements, and those covered by firm-level agreements.[4] They also report average contractual weekly working hours for full-time employees, as well as the occurrence of overtime, shift work, and weekend work. These measures allow us to capture both standard contractual hours and broader aspects of working-time organisation. The panel structure additionally enables us to examine within-establishment changes in working time following an exit from collective bargaining coverage. The dataset provides a rich set of establishment characteristics, including workforce composition, works council presence and establishment size, which we use as controls for potential confounding factors.

We impose several sample restrictions. First, we restrict the observation period to 2008 to 2024, since working-time information has been collected annually only from 2008 onwards and is available only for selected years before then. Second, for our main dependent variable, average contractual weekly working hours for full-time employees, we exclude implausible values to improve comparability. Specifically, we assume a standard five-day working week with a maximum of eight hours per day and exclude observations reporting more than 40 contractual hours per week. Although German labour law permits working time of up to 48 hours per week on the basis of a six-day working week, the five-day week remains the prevailing standard in both covered and uncovered establishments (Backhaus et al., 2022). Third, following previous empirical studies on working-time measurement, we also exclude establishments reporting average contractual weekly working hours for full-time employees below 35 hours (see Rengers et al., 2017; Schäfer, 2018). This lower threshold is motivated by the 35-hour week as a longstanding union benchmark, which is institutionalised in some sectors, such as metalworking, but is not universal (Bispinck & Schulten, 2017). Together, these restrictions exclude approximately the bottom and top five per cent of the contractual-hours distribution. We regard this as an appropriate trade-off: excluding implausible survey responses and potential outliers improves comparability while reducing the sample only moderately. As a robustness check, we relax the upper-bound restriction by allowing for a six-day working week and setting the maximum at 48 hours.

We apply additional restrictions to our key explanatory variable, collective bargaining coverage. First, where information on collective bargaining coverage is missing at the beginning or at the end of an establishment's observed panel history, we drop the respective annual observations but retain the establishment in the sample. Second, we exclude establishments entirely if information on collective bargaining coverage is missing within the retained observation period. This restriction avoids imputing or inferring changes in collective bargaining status when the timing of entry into, or exit from, coverage cannot be determined reliably. The resulting sample therefore consists of establishments with complete information on collective bargaining status for all retained years.

[4] In our observation period, the IAB Establishment Panel contains the information whether establishments are orientating towards a sectoral collective agreement with respect to wages and salaries. Given the specific wording this information cannot be used in our analysis.

We further exclude establishments that switch repeatedly between covered and uncovered status, as well as establishments that enter collective bargaining coverage during the observation period. Our analysis therefore focuses on establishments that exit coverage exactly once, which defines the treatment group and provides a clearly identifiable change in coverage status. This restriction reflects our interest in exits from collective bargaining and ensures that the within-establishment variation underlying the treatment indicator has a clear direction. Since fixed-effects estimations identify associations from within-establishment changes over time, focusing on single exits ensures that the estimated coefficient is based on a well-defined and interpretable source of variation.

Finally, we exclude establishments in the public and third sectors, following the classification proposed by Hohendanner et al. (2015) and in line with previous work (e.g. Bossler, 2019; Rieder & Schnabel, 2026). This restriction is motivated by institutional differences. In the public sector, working hours are largely shaped by statutory provisions and administrative regulations rather than by decentralised collective bargaining. The third sector includes church-affiliated establishments that are subject to a special labour-law regime in Germany. In these establishments, working conditions are not negotiated through collective bargaining between trade unions and employers, but are determined by parity-based labour-law commissions (Kreß, 2012). Excluding these sectors therefore yields a more homogeneous institutional setting for our analysis.

The control variables are chosen in line with the existing literature (Bossler, 2019; Rieder & Schnabel, 2026). We construct a dummy variable for the presence of a works council that equals one if a works council is present and zero otherwise. To reduce concerns about reverse causality, we use its lagged value in the empirical analyses. Workforce composition is captured by the shares of female, fixed-term, qualified, and part-time employees. The establishment's financial performance in the previous year is measured using a categorical variable indicating whether profits were positive, negative, or approximately balanced, or whether the establishment was unable to provide this information. We further construct a categorical variable for the establishment's management structure, distinguishing between establishments managed by an employed manager, by the owner, or by both. The quality of the establishment's technical equipment is captured by a dummy variable indicating whether the equipment is modern. We also construct a dummy variable equal to one if the establishment used overtime in the previous year and zero otherwise. This measure is used because exact data on the amount of overtime are not collected in every wave of the IAB Establishment Panel. Finally, establishment size is included as a categorical variable. These controls and the sample restrictions described above define the baseline sample used in our main analysis.

Building on this baseline sample, we construct a second dataset to examine dynamic adjustments around exits from collective bargaining coverage, following the logic of stacked event-study designs (Cengiz et al., 2019). We treat each exit from collective bargaining coverage as an event and examine changes in working time before and after exit within a difference-in-differences framework. Given the possibility of self-selection into exit, the estimates should be interpreted as descriptive associations rather than causal effects. The event-study approach is nevertheless useful for assessing whether working-time adjustments occur immediately or unfold gradually over time. This is particularly relevant in the German context, where collectively agreed provisions may continue to apply after an establishment leaves collective bargaining coverage until new employment contracts are concluded.

For establishments exiting collective bargaining coverage, the actual exit year defines the event year. Establishments that do not change coverage status serve as comparison units. To align them with treated establishments, we duplicate their observations across exit cohorts and assign artificial event years, corresponding to the observed exit years among treated establishments. This enables us to define event time consistently across treated and control group, allowing us to estimate our event-study.

## 5 Descriptive evidence

Figures 1 and 2 show the evolution of contractual weekly working hours for full-time employees by bargaining status and region. The latter is motivated by the suggestion of Oberfichtner and Schnabel (2019) as well as Schnabel (2016) to analyse both parts of Germany separately, since they differ fundamentally in terms of collective bargaining coverage and the presence of working councils, the two main pillars of the German industrial relations model.

Across the full observation period, uncovered establishments consistently report longer contractual working hours than covered establishments in both West and East Germany. In West Germany, working hours in uncovered establishments remain broadly stable, declining only slightly from 39.4 hours in 2008 to 39.2 hours in 2024. In East Germany, uncovered establishments initially report longer hours, averaging 39.8 hours in 2008, but these gradually converge towards West German levels, reaching 39.3 hours by 2024.

Among covered establishments in West Germany, contractual working hours vary little over time. Sectoral agreements are associated with stable working hours of around 39 hours. Firm-level agreements display somewhat greater variation, fluctuating between 38.3 and 39.2 hours and averaging 38.7 hours in 2024. This variation likely reflects small sample sizes, as firm-level agreements cover fewer than 2 per cent of establishments. In East Germany, differences between agreement types are comparatively small. In 2008, contractual working hours averaged 39.6 hours under firm-level agreements and 39.4 hours under sectoral agreements. While both groups show a downward trend over time, the decline is more pronounced for firm-level agreements, where average contractual working hours fall to 38.8 hours in 2024, whereas sectoral agreements remain comparatively stable, decreasing only slightly to around 39.1 hours over the same period.

These patterns are broadly consistent with other data sources. For 2024, the Federal Statistical Office reports average collectively agreed weekly hours of 38.5 in West Germany and 39.1 in East Germany[5], while the WSI Collective Agreement Archive[6] reports slightly lower levels of 37.6 and 38.6 hours, respectively.

(Figure 1 and Figure 2 around here)

Table1 reports average coverage rates and contractual weekly working hours over the full observation period for West and East Germany. For this comparison, establishments covered by sectoral and firm-level agreements are combined into a single category of covered establishments. Overall, coverage remains higher in West Germany than in East Germany, and covered establishments report shorter working hours than uncovered establishments in both parts of the country. We present sector specific results for West an East Germany in the online Appendix (see Tables OA1 –OA3).

---

[5] Federal Statistical Office (Destatis), GENESIS-Online, Table 62221-0003, https://genesis.destatis.de/datenbank/online/table/62221-0003.

[6] WSI-Tarifarchiv, “Tarifstatistik: Wochenarbeitszeit,” https://www.wsi.de/de/wochenarbeitszeit-15326.htm.

(Table 1 around here)

We then move beyond cross-sectional differences between covered and uncovered establishments and examine working-time changes around exits from collective bargaining coverage by comparing average weekly working hours in the years before and after exit. Table 2 shows that exits are associated with increases in contractual working hours in both West and East Germany. In West Germany, establishments that exit coverage report an average of 39.07 hours before exit. This is close to the average for permanently uncovered establishments, at 39.38 hours, and slightly above that for covered establishments, at 38.96 hours. After exit, contractual working hours increase modestly to 39.19 hours. Although this increase is statistically significant, it amounts to only about 7 minutes per week and is therefore economically small.

A similar pattern is observed in East Germany. Before exit, establishments report average working hours of 39.47 hours, close to the average among permanently uncovered establishments, at 39.58 hours. Following exit, contractual working hours increase significantly, with the increase amounting to approximately 9 minutes per week. The effect is therefore again economically small.

(Table 2 around here)

## 6 Empirical model

To examine the relationship between collective bargaining coverage and establishment-level working time, we estimate both pooled ordinary least squares (POLS) and fixed-effects (FE) models. The aim of using POLS is primarily to relate our findings to the existing literature and should be seen more as descriptive evidence. The reason is that the POLS specification is driven by between-establishment differences, whereas the FE model exploits within-establishment variation and controls for unobserved time-invariant heterogeneity leading to more unbiased results (Cameron & Trivedi, 2022; Wooldridge, 2010). Since the FE estimator relies on changes within establishments over time, it is our preferred specification for analysing how establishments adjust working time after exiting collective bargaining coverage. To ensure that the identifying variation reflects exits from coverage, we exclude establishments that enter coverage during the observation period, so that variation in the coverage indicator is one-sided.

Building on this framework, we extend the analysis using a difference-in-differences event-study design based on a second, stacked dataset, treating each exit from collective bargaining coverage as a discrete-time event. In contrast to conventional event-study designs that aim to identify causal effects, our estimates should be interpreted as descriptive associations because we cannot account fully for establishments' self-selection into collective bargaining exit. Nevertheless, this approach allows us to examine the timing and persistence of working-time adjustments, including whether such adjustments are delayed by the continued applicability of collectively agreed provisions after exit. We conduct the analysis separately for West and East Germany to account for persistent differences in collective bargaining coverage and other labour-market characteristics (Oberfichtner & Schnabel, 2019; Schnabel, 2016).

Our baseline specification is:

$$y_{it} = \beta nca_{it} + x_{it} + \alpha_{fs} + \delta_c + \gamma_s + \phi_e + \epsilon_{i,t}$$

Here, $y_{it}$denotes the logarithm of working time in establishment $i$ at time $t$, and $nca_{it}$ indicates non-coverage by a collective agreement. The dummy takes the value of one when the establishment is not covered by an agreement and zero otherwise. The vector $x_{it}$ includes time-varying establishment characteristics, namely lagged information on the presence of a works council, workforce composition (the shares of female, fixed-term, qualified, and part-time employees), financial performance in the previous year, management structure, technical equipment, overtime in the previous year, and establishment size.

All specifications include calendar-year fixed effects ($\delta_c$). The POLS models include federal-state ($\alpha_{fs}$) and sector fixed effects ($\gamma_s$), while the FE models additionally include establishment fixed effects ($\phi_e$)., which absorb time-invariant establishment characteristics, including federal state and sector. The term $\epsilon_{i,t}$ represents the idiosyncratic error term. In all models, standard errors are clustered at the establishment level.

For the event-study specification, we estimate:

$$y_{i,t} = \sum_{\tau \neq -1} \beta_\tau I(t = \tau) + \sum_{\tau \neq -1} \gamma_\tau I(t = \tau) I(T_i = 1) + x_{it} + \alpha_{fs} + \delta_c + \gamma_s + \phi_e + \epsilon_{i,t}$$

where $T_i$ is a treatment indicator equal to one if an establishment exits collective bargaining coverage and zero otherwise. For a given event time $\tau$, $\beta_\tau$ captures the average change in the outcome variable for the control group (establishments that never change coverage status) relative to the reference period, while $\gamma_\tau$ measures the difference between treated and control establishments at that time. Event time $\tau$ ranges from −16 to +15. To avoid collinearity, periods with $\tau \leq -6$ or $\tau \geq 7$ are grouped as suggested by Miller (2023), with $\tau = -1$ as the reference period.

## 7 Results

We begin by reporting pooled OLS estimates that document differences in contractual weekly working hours between establishments covered by collective bargaining and those without coverage in West and East Germany. The purpose of this exercise is primarily descriptive: it allows us to relate our findings to the existing literature on working-time differences by bargaining status.

Table 3 reports the pooled OLS estimates together with sample means of average agreed weekly working hours for covered and uncovered establishments.[7] In both West and East Germany, establishments without collective bargaining coverage display statistically significantly longer working hours than covered establishments. Conditional on the included controls, uncovered establishments in West Germany have, on average, 2.10 per cent longer contractual weekly working hours than covered establishments. In East Germany, the corresponding difference amounts to around 1.33 per cent.

Expressed in time units, these estimates imply differences in agreed weekly working hours. Based on the average working time of 38.30 hours in covered establishments in West Germany and 39.13 hours in East Germany, the estimated difference corresponds to approximately 48 minutes per week in West Germany and around 31 minutes per week in East Germany.[8] The larger difference in West Germany is consistent with the descriptive patterns reported in Table 1, although the differences are more pronounced in the multivariate analysis.

[7] For all tables, coefficients on control variables are available upon request.

Overall, the pooled OLS results confirm the finding of our descriptive results that collective bargaining coverage is associated with shorter agreed weekly working hours.

(Table 3 around here)

We next move beyond differences between covered and uncovered establishments and examine how exits from collective bargaining coverage are associated with changes in contractual working hours. For this purpose, we estimate fixed-effects models. Unlike the pooled OLS estimates, which are primarily driven by between-establishment variation and therefore mainly capture differences between establishments with and without collective bargaining coverage, the fixed-effects specification exploits within-establishment variation and controls for unobserved time-invariant establishment characteristics. Since establishments entering collective bargaining coverage are excluded from the analysis, the identifying variation results exclusively from establishments that exit coverage during the observation period. The results are reported in Table 4.

The fixed-effects estimates are considerably smaller than the pooled OLS estimates. In West Germany, establishments leaving collective bargaining coverage increase contractual weekly working hours by approximately 0.33 per cent. Although the estimate is statistically significant at the one per cent level, its economic magnitude is small. Evaluated at the mean weekly working time of covered establishments, 38.30 hours (see Table 4), the estimate corresponds to an increase of only about eight minutes per week after exit. A similar pattern emerges in East Germany, where working time increases by approximately 0.32 per cent after exit from collective bargaining coverage. This estimate is also statistically significant at the one per cent level, but also corresponds to only about eight additional minutes per week, based on an average weekly working time of 39.13 hours in covered establishments.

While these weekly changes are economically modest, simple back-of-the-envelope calculations suggest that they are not entirely negligible when accumulated over the year.

[8] Minute equivalents are obtained by multiplying the mean weekly working hours reported in Table 3 of covered establishments by 60 and by the estimated coefficient.

An increase of around eight minutes per week corresponds to roughly seven additional hours per year, or almost one additional working day. [9] Sector specific analysis for pooled OLS and fixed effects estimators are presented in the online appendix see Figures OA1 and OA2.

(Table 4 around here)

While the average fixed-effects estimates capture the overall association with leaving collective bargaining coverage, they do not reveal when establishments adjust working time after the transition. To examine the timing of these adjustments, we use our second dataset and estimate event-study models separately for West and East Germany. These two-way fixed-effects models compare establishments that leave collective bargaining coverage with establishments whose bargaining status remains unchanged throughout the observation period.

This is particularly relevant in the German context, as collective agreement provisions continue to apply after an establishment exits collective bargaining coverage until they are replaced by a new agreement. The event-study estimates therefore allow us to examine whether establishments leaving collective bargaining coverage already differed in the pre-period before exit and to determine the timing of any adjustments in working hours during the post-exit period.

(Figure 3 around here)

Figure 3 reports the results for West Germany in blue and East Germany in red. Turning first to West Germany, the estimates show that establishments leaving collective bargaining coverage had, on average, slightly shorter working hours than establishments whose bargaining status did not change during the five years before exit. However, these differences are not statistically significant. Thus, our results suggest that working hours followed similar trends in establishments that left collective bargaining coverage and those whose coverage status remained unchanged. From the year of exit onwards, average weekly working hours among full-time employees increase. In the first year after exit, the difference amounts to around 0.42 per cent and is statistically significant. This positive association persists until the fourth year after exit, although the estimate in year four is no longer statistically significant and amounts to 0.34 per cent. In the fifth and sixth years after exit, the estimated differences decline to 0.09 and 0.13 per cent, respectively. By contrast, the long-run estimate for seven or more years after exit is again statistically significant and indicates an increase in working time of around 0.47 per cent.

[9] Annual equivalents are calculated as a simple yearly approximation by multiplying the estimated weekly increase in minutes by 52 weeks and dividing the result by 60.

These results can be interpreted as evidence that, in West Germany, working-time adjustments take place relatively soon after establishments leave collective bargaining coverage and this is associated with higher working time in the longer run. However, given that the average weekly working time of covered establishments in West Germany is 38.97 hours (see Table 1), this increase corresponds to only around 11 minutes per week. The economic magnitude is therefore limited.

A similar pattern emerges in East Germany. Apart from the fourth year before exit, there are no statistically significant differences in working hours between establishments leaving collective bargaining coverage and establishments remaining in the same bargaining-status group before exit. After exit from collective bargaining coverage, working time increases in exiting establishments relative to establishments without a status change. In the first year after exit, the increase amounts to 0.29 per cent. The estimates remain positive in subsequent years and reach around 0.5 per cent by the sixth year after exit. The long-run estimate for seven or more years after exit is also of a similar magnitude, at around 0.6 per cent.

Nevertheless, as in West Germany, it is questionable how economically relevant these effects are. Even the sixth-year estimate for East Germany corresponds to an increase of only around 12 minutes per week, evaluated at the average weekly working time of covered establishments of 39.24 hours. At the same time, the event-study estimates for East Germany, as in West Germany, suggest that establishments adjust working time relatively quickly after leaving collective bargaining coverage. This raises the policy-relevant question of how much protection the legally regulated after-effect of collective agreements provides for employees if establishments already adjust working-time arrangements shortly after exiting coverage. However, this is a question that cannot be fully answered in this paper.

## 8 Robustness checks

We assess the robustness of our main results in four ways. First, we relax the upper bound of 40 hours and restrict the sample by excluding only establishments that report agreed average weekly working hours for full-time employees above the statutory maximum of 48 hours. Second, our observation period includes the Covid-19 pandemic, which may have affected the reporting behaviour of surveyed establishments. For example, if working hours were reduced through short-time work, respondents may have incorporated this reduction into the reported average, even though this should not directly affect contractually agreed working hours. We address this concern by excluding the period from 2020 to 2024 from the POLS and fixed-effects estimations. In the event-study specification, we exclude establishments that exited collective bargaining coverage in 2020 or later.

Third, while following the literature by choosing our control variables, we are concerned that some of our control variables might be reversed influence by our outcome variable. For example, one could argue that the share of women working in an establishment might influence working hours. At the same time, it is also possible to argue that fewer women join or leave the establishment because they are put off by the required flexibility due to long working hours. This would lead to biased results of reasons such as bad controls or reverse causality problem. To address this, we repeat the analysis using only time invariant control variables such as federal-state, industry and year dummies as well as establishment fixed effects in our analysis. If our results remain similar to those of our main specification, we conclude that potential bad-control concerns play only a minor role in our setting. A detailed discussion of the robustness checks as well as our results is provided in the online appendix (see Tables OA4–OA8 and Figures OA3–OA9).

Tables OA4 and OA5 in the appendix report the corresponding estimates when the upper bound for agreed average weekly working hours of full-time employees is extended to 48 hours. Figures OA3 and OA4 additionally present the event-study results for both West and East Germany. Results from the POLS and fixed-effects estimations for both parts of the country are very similar to those in the main specification. With regard to the POLS estimates, the significance levels remain unchanged and the point estimates are almost identical. The same also holds for the fixed-effects estimates for West Germany. The only exception is the fixed-effects estimate for East Germany, where the coefficient is no longer statistically significant. However, with a coefficient of about 0.2 per cent, it still indicates a comparable association to that in the main specification, where the corresponding estimate is approximately 0.3 per cent. We therefore consider the POLS and fixed-effects results to be robust. The event-study estimates (Figures OA3 and OA4) show a similar pattern, although some coefficients lose statistical significance in both regions. However, as the confidence intervals overlap substantially and the point estimates are similar in magnitude to those of the main specification, we interpret the results as robust.

Excluding the years from 2020 onwards does not materially affect the POLS or fixed-effects estimates for West and East Germany. The POLS estimates (Table OA6) are nearly identical in both parts of Germany with respect to both the size of the coefficients and their statistical significance. The fixed-effects estimates (Table OA7) for West and East Germany remain statistically significant, although now at the five per cent level rather than the one per cent level. For both parts of Germany, the point estimates are somewhat smaller but still comparable to those in the main specification. The event-study estimates (Figures OA5 and OA6) show a broadly similar overall pattern to the main specification, although some coefficients lose statistical significance. We therefore conclude that our findings are not driven by the Covid-19 pandemic.

When considering only federal-state, industry, and year fixed effects in our POLS and fixed effects estimates (Table OA8), the results remain similar in terms of both statistical significance and the magnitude of the point estimates, although the differences in the coefficients are slightly larger in West Germany. The results from the event-study results (Figures OA7 and OA8) are also consistent across both specifications. We therefore conclude that bad controls do not appear to pose a problem in our estimations.

Fourth, dynamic two-way fixed effects (TWFE) estimators can yield misleading estimates when treatment timing is staggered and effects vary across cohorts. Comparisons involving already-treated units can generate negative weights and contaminate event-time coefficients with effects from other periods. This may produce estimates with the wrong sign and apparent pre-trends even when parallel trends hold (Roth et al., 2023; Sun & Abraham, 2021). Our stacked event-study design compares each exit cohort exclusively with establishments that never change their bargaining status, thereby avoiding comparisons in which already-treated establishments serve as controls. As an additional robustness check, we apply the estimator proposed by Callaway and Sant'Anna (2021), which accommodates heterogeneous treatment effects, to Germany as a whole and separately to western and eastern Germany.

Unlike in the baseline specification, we do not bin the endpoints when using the Callaway and Sant'Anna (2021) estimator. The estimates therefore refer specifically to six years before and seven years after exit, rather than six or more years before and seven or more years after exit. The results, reported in Figure OA9 in the online appendix, provide no evidence of differential pre-trends in any of the three samples. Post-exit patterns in western and eastern Germany are similar to those in the baseline specification, supporting the robustness of our findings.

## 9 Other margins

In addition to contractual working hours, the IAB Establishment Panel collected biennial information on the use of working-time organisation and adjustment instruments until 2020. Establishments were asked whether they used specific instruments, with binary response categories. These instruments include regular Saturday and Sunday work, demand-dependent Saturday work, shift work, staggered working hours, adjustments to part-time working hours, working-time corridors or bandwidth models, and working-time reductions aimed at safeguarding employment.

Using this information for our observation period from 2008 onwards, we construct binary indicators for each category and estimate separate linear probability models. As before, we report pooled ordinary least squares and fixed-effects estimates. The pooled OLS estimates capture average differences between establishments with and without collective bargaining coverage, whereas the fixed-effects estimates focus on changes associated with leaving collective bargaining coverage.

(Figure 4 around here)

Figure 4 shows the results of our POLS estimations for West and East Germany on the left-hand side and the results of the fixed-effects estimations on the right-hand side. The POLS results indicate that, ceteris paribus, in both West and East Germany, establishments uncovered by collective agreements are significantly less likely than covered establishments to use regular Saturday work, staggered working hours and working-time corridors. However, only in West Germany do the estimates indicate that uncovered establishments are also significantly less likely to use shift work.

Regarding the effects of exits on working-time arrangements, the fixed-effects estimates show that in both parts of Germany, an exit from collective bargaining coverage is associated with a higher probability of using Sundays as regular working days and a lower probability of using Saturday as regular working days in response to operational needs. But only in East Germany the exit is associated with a higher probability of using Saturdays as regular working days. In both regions, the point estimates further indicate a lower use of working-time corridors.

Taken together, the fixed-effects estimates suggest only small changes in working-time arrangements, with mixed signs across mechanisms. While some point estimates indicate greater use of arrangements that may have been more costly under collective agreements, others point in the opposite direction. These findings should be interpreted as suggestive only. None of the fixed-effects estimates is statistically significantly different from zero, and the wide confidence intervals indicate substantial statistical uncertainty. This imprecision likely reflects both the small number of establishments leaving collective bargaining coverage and the fact that the relevant working-time measures are not collected annually in the IAB Establishment Panel.

## 10 Conclusion

Our study provides novel evidence from the IAB Establishment Panel on how contractually agreed weekly working hours change when establishments leave collective bargaining coverage. Consistent with previous research (e.g. Bechmann et al., 2015; Kölling & Lehmann, 2002; Lübker & Schulten, 2024) we find that agreed working hours are higher in East than in West Germany and lower in establishments covered by collective agreements than in uncovered establishments. However, the adjustments of working hours after leaving collective bargaining coverage are modest. In both East and West Germany, the fixed-effects estimates indicate an increase of approximately 0.3 per cent in agreed weekly working hours for full-time employees, corresponding to around eight additional minutes per week, while the event-study estimates indicate a long-term increase of around 0.5 to 0.6 per cent, corresponding to around eleven to thirteen minutes per week. Moreover, the event-study estimates indicate that the increase emerges shortly after exits from collective bargaining coverage and persists over time. This finding is unlikely to be driven mechanically by post-contractual continuation obligations. Translating our weekly estimates into annual terms, exiting collective bargaining coverage is associated with approximately one additional working day per year.

These findings indicate that the additional formal autonomy gained by leaving collective bargaining coverage translates into only limited adjustments in contractual weekly hours.

If establishments exited primarily to increase working hours, one would expect them to make substantially greater use of this additional discretion. Instead, we observe only modest increases in agreed hours and no statistically significant changes in the other working-time arrangements covered by our data, including Saturday and Sunday work. Therefore, it is questionable whether working time regulations are one of the main drivers to leave collective bargaining coverage as found in surveys by Langer (1994), Lehmann (2002) and Lesch et al. (2018). At the same time, it might be possible that employers cannot freely adjust working hours against their employees' preferences, even if they have opted out of collective agreements. This could prompt employees to seek employment elsewhere.

Our analysis is subject to several limitations. First, the estimates capture within-establishment associations rather than causal effects, as establishments self-select into leaving collective bargaining coverage. Second, the data measure average agreed weekly working hours at the establishment level rather than actual hours worked by individual employees. Establishments may therefore adjust labour input through overtime, shift schedules, the distribution of hours across the week, weekend work, or heterogeneous changes within the workforce. Although our analysis of these margins provides little evidence of substantial changes in observable working-time arrangements, we cannot rule out adjustments in areas not covered by the data. Third, our main outcome captures the establishment-level average of contractually agreed weekly working hours among full-time employees. The survey provides no comparable information for part-time employees, and it remains unclear whether or how they are considered when establishments report this average. But this should be just a minor issue in our analysis. Our estimates are unbiased as long as reported working hours are not systematically under- or over reported. The establishment-level measure may therefore conceal substantial heterogeneity in working-time adjustments across individual employees and employment groups. This limitation may be particularly relevant because survey evidence suggests that part-time employees are more likely than full-time employees to wish to increase their working hours. We therefore cannot determine how adjustments following exits from collective bargaining coverage are distributed across the workforce.

Overall, our findings provide only limited support for the view that establishments leave collective bargaining coverage in order to substantially increase contractual working hours. They are in line with recent evidence from the IAB Vacancy Survey showing that the number of establishments reporting a need for employees to work more than ten hours on individual days is substantially smaller when the analysis is restricted to collectively covered establishments than across the economy as a whole. Although this pattern varies considerably across establishment sizes and industries, it suggests that collectively covered establishments account for only a limited share of the overall demand for this form of working-time flexibility (Hellwagner et al., 2026). While leaving collective bargaining coverage increases establishments' formal scope for unilateral decision-making, employers make only limited use of this additional autonomy with respect to contractual weekly hours. Taken together, our results show, in line with the OECD (2019, 2021), that working time remains an important subject of collective bargaining, but suggest that constraints on contractual working hours may be less binding in practice than employer reports imply.

## Tables

**Table 1. Collective bargaining coverage and agreed weekly working hours of full-time employees in West and East Germany**

| | (1) | (2) | (3) | (4) | | |
|---|---|---|---|---|---|---|
| | N | CBA coverage (%) | Hours with CBA | Hours without CBA | Difference (4) - (3) | Diff. in minutes |
| West Germany | 88,593 | 24.85 | 38.97 | 39.37 | 0.40*** | 24.0 |
| East Germany | 63,010 | 13.28 | 39.24 | 39.58 | 0.34*** | 20.4 |

Notes: The table reports collective bargaining agreement (CBA) coverage and average agreed weekly working hours for full-time employees by region. CBA coverage indicates whether the establishment is covered by a collective bargaining agreement. N denotes unweighted establishment-year observations used to calculate the weighted averages; all percentages, means, and differences are weighted. The difference is calculated by subtracting weekly hours in establishments with CBA coverage from weekly hours in establishments without CBA coverage and is converted into minutes by multiplying by 60. Statistical significance is based on t-tests of the difference between hours with and without CBA coverage using Stata's lincom command. * p < .10, ** p < .05, *** p < .01. Source: IAB-Establishment Panel, waves 2008-2024.

**Table 2. Agreed weekly working hours before and after exiting CBA coverage in West and East Germany**

| | (1) | (2) | (3) | (4) | | |
|---|---|---|---|---|---|---|
| | Always CBA | Never CBA | Before exit | After exit | Difference (4) - (3) | Diff. in minutes |
| West Germany | 38.96 | 39.38 | 39.07 | 39.19 | 0.12*** | 7.2 |
| N | 28,303 | 53,002 | 3,079 | 4,209 | | |
| N-Establishments | 8,796 | 18,141 | 1,265 | 1,228 | | |
| East Germany | 39.19 | 39.58 | 39.47 | 39.63 | 0.15*** | 9.0 |
| N | 10,539 | 48,177 | 1,591 | 2,703 | | |
| N-Establishments | 3,218 | 13,806 | 695 | 676 | | |

Notes: The table reports average agreed weekly working hours of full-time employees in Germany by region and CBA coverage status. “Always CBA” and “Never CBA” denote establishments continuously covered and never covered by a CBA, respectively; “Before exit” and “After exit” refer to observations before and after leaving CBA coverage. N denotes unweighted establishment-year observations and N-Establishments the number of unique establishments used to calculate the weighted averages; all means and differences are weighted. Differences are calculated as hours without CBA minus hours with CBA and converted into minutes by multiplying by 60. Statistical significance is based on t-tests using Stata’s lincom command. * $p < .10$, ** $p < .05$, *** $p < .01$. Source: IAB-Establishment Panel, waves 2008-2024.

**Table 3. Pooled OLS estimates of non-collective bargaining coverage and agreed weekly working time among full-time employees in West and East Germany**

| | West | East |
|---|---|---|
| Non-CBA coverage (yes = 1; no =0) | 0.0210*** (0.0008) | 0.0133*** (0.0010) |
| Year FE | yes | yes |
| Industry FE | yes | yes |
| Federal-state FE | yes | yes |
| Establishment FE | no | no |
| Adjusted $R^2$ | 0.248 | 0.090 |
| Mean, CBA | 38.30 | 39.13 |
| Mean, non-CBA | 39.38 | 39.68 |
| Observations | 57,267 | 44,180 |
| CBA observations | 19,438 | 7,634 |
| Non-CBA observations | 37,829 | 36,546 |

Notes: The table reports pooled ordinary least squares estimates for West and East Germany. The dependent variable is the logarithm of agreed weekly working time for full-time employees. Collective bargaining agreement (CBA) coverage indicates whether the establishment is not covered by a collective bargaining agreement. Standard errors, clustered at the establishment level, are reported in parentheses. All models include year fixed effects, federal-state fixed effects, industry fixed effects, establishment size controls, workforce composition controls, and additional establishment-level controls. Workforce composition controls include the shares of female, fixed-term, qualified, and part-time employees. Establishment-level controls include lagged works council presence, profit situation in the previous year, management structure, state of technical equipment, and an indicator for overtime in the previous year. * p < .10, ** p < .05, *** p < .01. Source: IAB-Establishment Panel, waves 2008-2024.

**Table 4. Fixed-effects estimates of exits from collective bargaining coverage and agreed weekly working time among full-time employees in West and East Germany**

| | West | East |
|---|---|---|
| Non-CBA coverage (yes = 1; no =0) | 0.0033***<br>(0.0012) | 0.0032***<br>(0.0011) |
| Year FE | yes | yes |
| Industry FE | yes | yes |
| Federal-state FE | yes | yes |
| Establishment FE | yes | yes |
| Adjusted $R^2$ | 0.002 | 0.012 |
| Mean, CBA est. | 38.30 | 39.13 |
| Mean, non-CBA est. | 39.38 | 39.68 |
| Observations | 57,267 | 44,180 |
| CBA observations | 19,438 | 7,634 |
| Non-CBA observations | 37,829 | 36,546 |

Notes: The table reports fixed-effects estimates for West and East Germany. The dependent variable is the logarithm of agreed weekly working time for full-time employees. Collective bargaining agreement (CBA) coverage indicates whether the establishment is not covered by a collective bargaining agreement. Establishments entering collective bargaining agreement coverage during the observation period are excluded; therefore, within-establishment variation in coverage reflects exits from coverage. As a result, the coefficient indicates an increase in agreed weekly working time after exiting collective bargaining agreement coverage. Standard errors, clustered at the establishment level, are reported in parentheses. All models include year fixed effects, establishment fixed effects, establishment size controls, workforce composition controls, and additional establishment-level controls. Workforce composition controls include the shares of female, fixed-term, qualified, and part-time employees. Establishment-level controls include lagged works council presence, profit situation in the previous year, management structure, state of technical equipment, and an indicator for overtime in the previous year. * $p < .10$, ** $p < .05$, *** $p < .01$. Source: IAB-Establishment Panel, waves 2008-2024.

## Figures

### Figure 1. Average weekly working hours among full-time employees by collective bargaining status in West Germany, 2008-2024

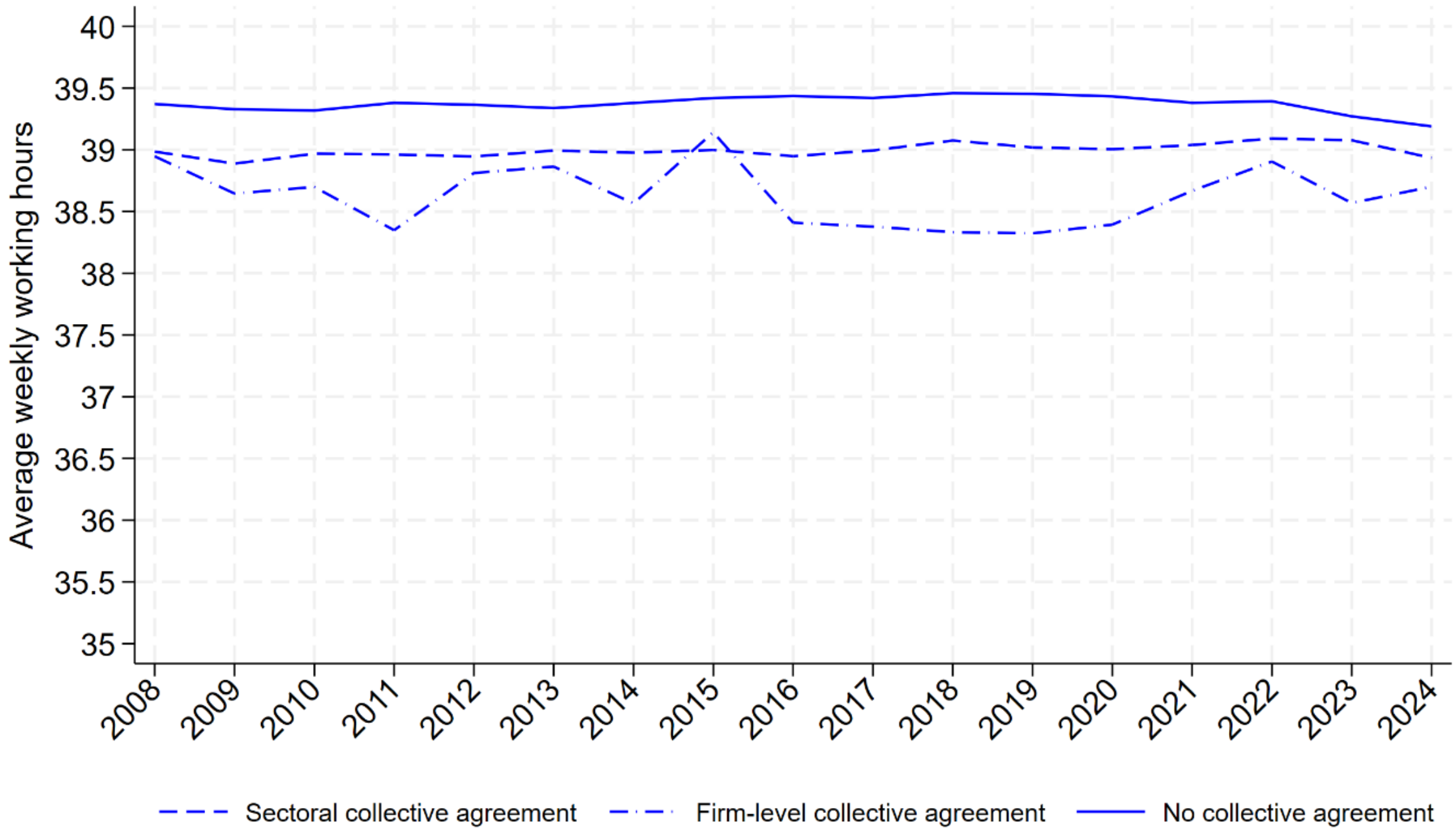

Notes: The figure reports average weekly working hours among full time employees per establishment, separately for West Germany and by collective bargaining status. Dashed lines show establishments covered by sectoral collective agreements; dash-dotted lines show establishments covered by firm-level collective agreements; solid lines show establishments not covered by a collective agreement. Estimates are weighted using the establishment weights from the IAB Establishment Panel. The sample excludes establishments in the third sector. Source: IAB-Establishment Panel, waves 2008-2024.

**Figure 2. Average weekly working hours among full-time employees by collective bargaining status in East Germany, 2008-2024**

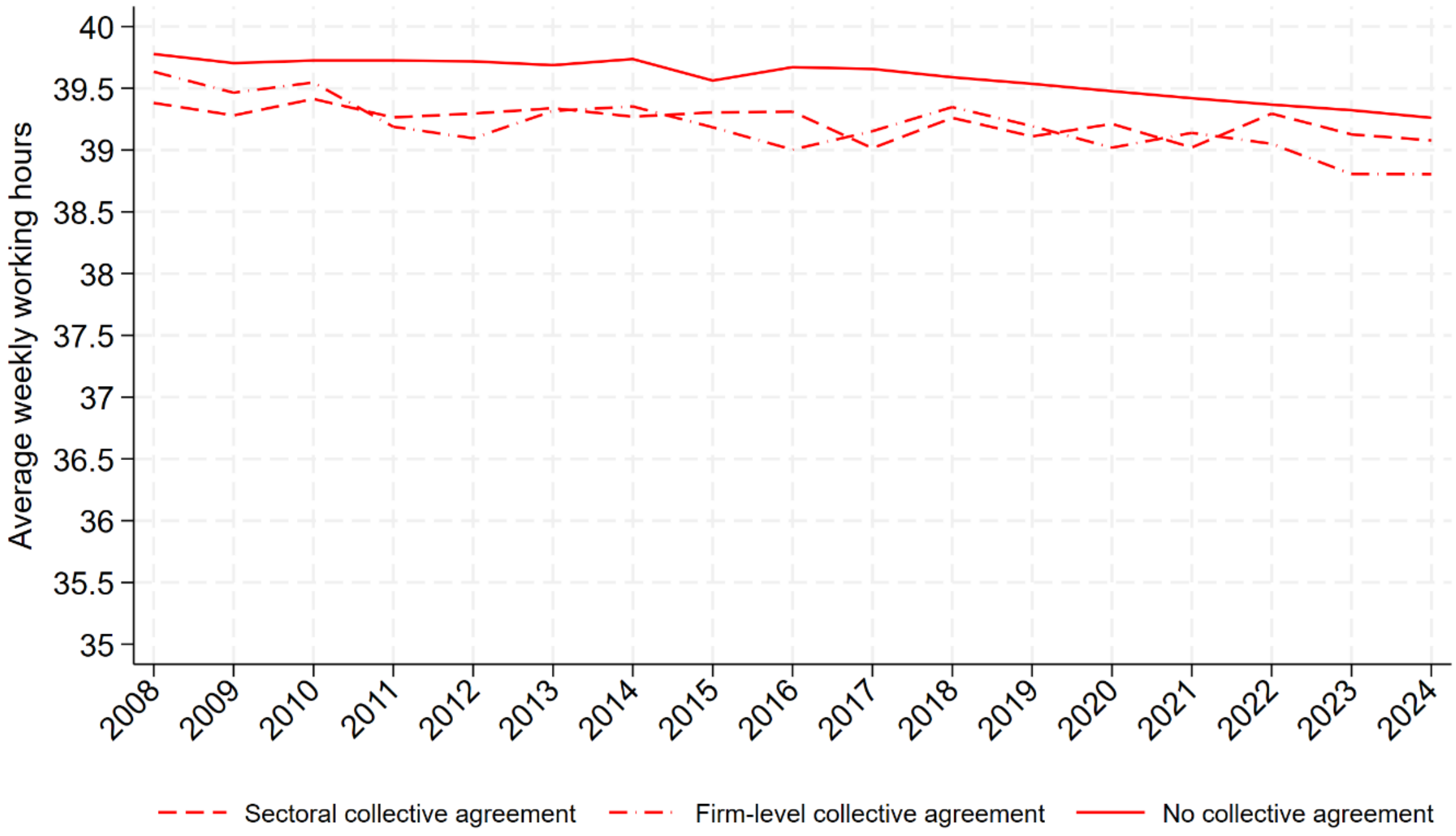

Notes: The figure reports average weekly working hours among full-time employees per establishment, separately for East Germany and by collective bargaining status. Dashed lines show establishments covered by sectoral collective agreements; dash-dotted lines show establishments covered by firm-level collective agreements; solid lines show establishments not covered by a collective agreement. Estimates are weighted using the establishment weights from the IAB Establishment Panel. The sample excludes establishments in the third sector. Source: IAB-Establishment Panel, waves 2008-2024.

**Figure 3. Event-study estimates of agreed weekly working time among full-time employees around exit from collective bargaining coverage in West and East Germany**

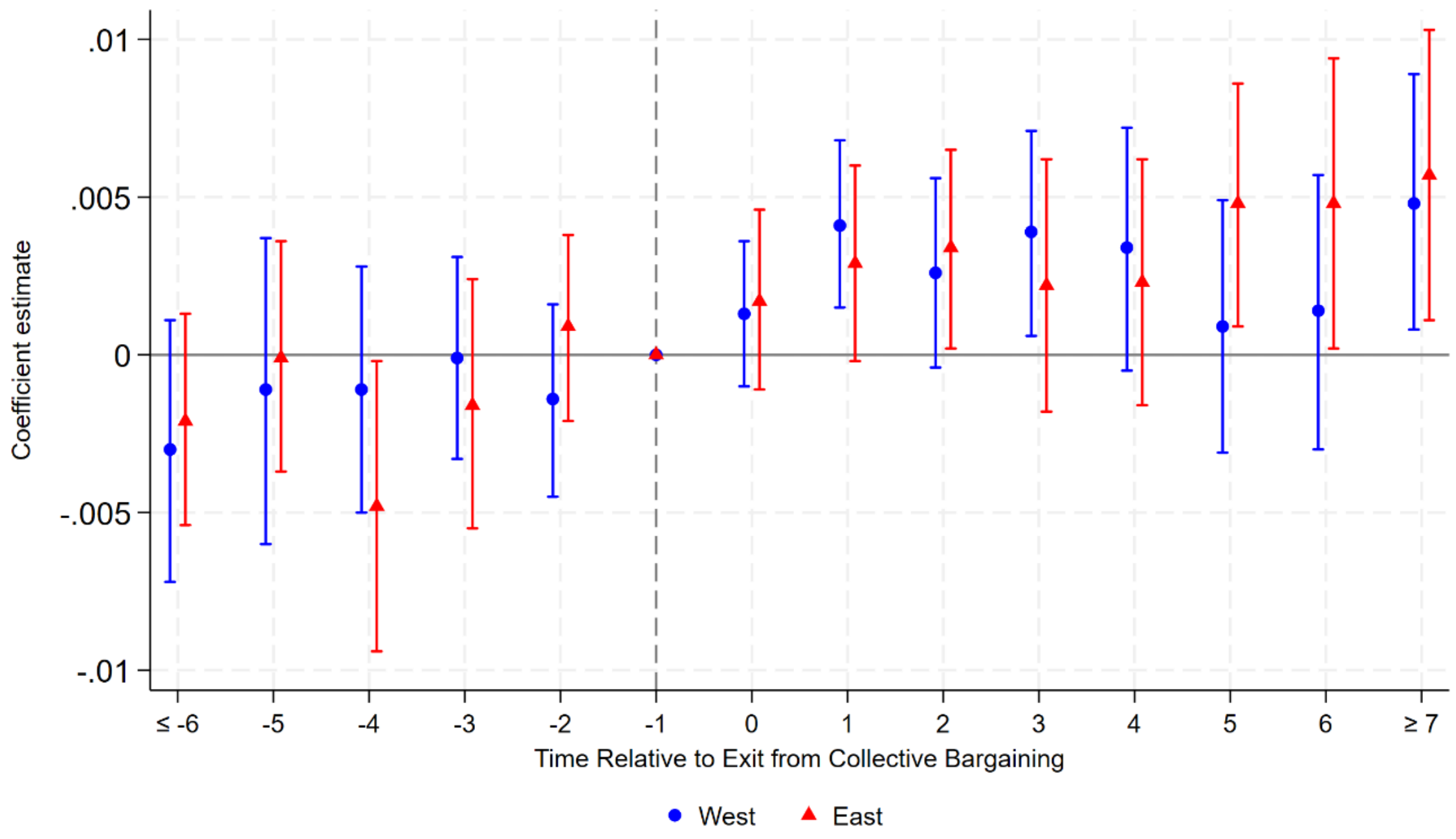

Notes: The figure reports event-study estimates of the change in agreed weekly working time around exit from collective bargaining coverage, separately for West and East Germany. The dependent variable is the logarithm of agreed weekly working time for full-time employees. Coefficients are measured relative to the year before exit from collective bargaining coverage. Points show coefficient estimates and vertical lines show 95% confidence intervals. The event-time categories ≤ -6 and ≥ 7 pool observations six or more years before and seven or more years after exit, respectively. Standard errors are clustered at the establishment level. Models include establishment fixed effects, year fixed effects, federal-state fixed effects, establishment size controls, workforce composition controls, and establishment-level controls. Workforce composition controls include the shares of female, fixed-term, qualified, and part-time employees; establishment-level controls include lagged works council presence, previous-year profit situation, management structure, state of technical equipment, and an indicator for overtime in the previous year. The estimation sample includes 778,001 establishment-year observations in West and 614,330 in East Germany. Source: IAB-Establishment Panel, waves 2008-2024.

**Figure 4. POLS and fixed-effects estimates of the use of alternative forms of working time adjustments around exit from collective bargaining coverage in West and East Germany**

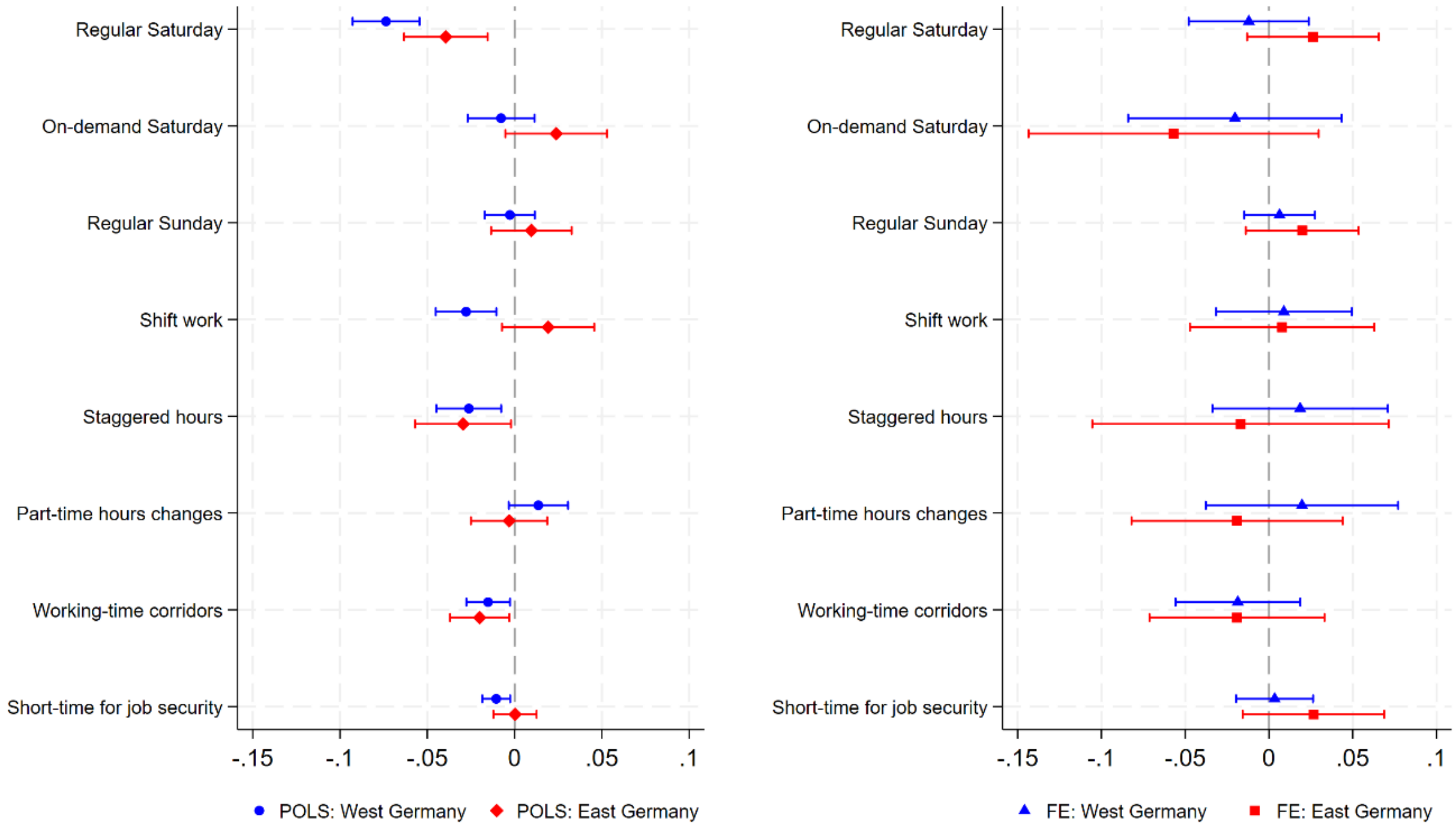

Notes: The figure reports pooled OLS (POLS) and fixed-effects (FE) estimates of alternative forms of working-time adjustment, separately for West and East Germany. Dependent variables are dummies indicating whether the respective working-time instrument is used. Coefficients and 95% confidence intervals are reported; standard errors are clustered at the establishment level. The POLS model includes year fixed effects, federal-state fixed effects, establishment size controls, workforce composition controls, and establishment-level controls while the FE model includes additionally establishment fixed effects. N denotes establishment-year observations and is identical for POLS and FE. In East Germany, N is 16,478 for regular Saturday work, 16,442 for Saturday work when required, 16,462 for Sunday work, 16,473 for shift work, 16,495 for staggered working hours, 16,462 for working-time adjustments for part-time employees, 16,444 for working-time corridors, and 16,431 for employment-safeguarding working-time reductions. In West Germany, N is 21,613 for regular Saturday work, 21,556 for Saturday work when required, 21,555 for Sunday work, 21,571 for shift work, 21,615 for staggered working hours, 21,556 for working-time adjustments for part-time employees, 21,508 for working-time corridors, and 21,477 for employment-safeguarding working-time reductions. Source: IAB-Establishment Panel, waves 2008 , 2010, 2012, 2014, 2016, 2018, 2020.

## Online Appendix

### *Sector analysis: Descriptives*

Industry-level comparisons reveal substantial heterogeneity in both bargaining coverage and working hours. Table OA1 reports average coverage rates and contractual weekly working hours by industry over the full observation period for West and East Germany. For this comparison, establishments covered by sectoral and firm-level agreements are combined into a single category of covered establishments.

In West Germany differences across industries are substantial. Collective bargaining coverage reaches 54.4 per cent in construction but only 11.8 per cent in transport, logistics, information and communication. Differences in working hours between covered and uncovered establishments also vary markedly across sectors. In construction, the gap is negligible, amounting to 2.4 minutes per week, and is statistically insignificant. By contrast, in financial, insurance and business services, the difference is both statistically significant and substantively larger, at 51.6 minutes per week. A similar pattern emerges in East Germany. Coverage is highest in construction, at approximately 38.5 per cent, but the associated difference in working hours remains modest at around 7 minutes, albeit statistically significant. In contrast, financial, insurance and business services combine low coverage, at around 12 per cent, with a substantially larger working-time differential of approximately one hour per week.

**Table OA1. Collective bargaining coverage and agreed weekly working hours of full-time employees by industry and region**

| West Germany | (1) | (2) | (3) | (4) | | |
|---|---|---|---|---|---|---|
| Industry | N | CBA coverage (%) | Hours with CBA | Hours without CBA | Difference (4) - (3) | Diff. in minutes |
| All sectors | 88,593 | 24.85 | 38.97 | 39.37 | 0.40*** | 24.0 |
| Manufacturing | 20,769 | 26.46 | 38.40 | 39.25 | 0.84*** | 50.4 |
| Agriculture, mining, utilities and waste | 2,848 | 31.73 | 39.49 | 39.69 | 0.20*** | 12.0 |
| Construction | 7,497 | 54.37 | 39.46 | 39.49 | 0.04 | 2.4 |
| Trade | 16,503 | 24.38 | 38.89 | 39.36 | 0.47*** | 28.2 |
| Transport, logistics, information and communication | 6,649 | 11.77 | 39.10 | 39.54 | 0.43*** | 25.8 |
| Hospitality, food and beverages | 6,843 | 27.48 | 39.28 | 39.51 | 0.23*** | 13.8 |
| Finance, insurance and business services | 17,170 | 13.14 | 38.60 | 39.45 | 0.86*** | 51.6 |
| Education, health and other services | 10,314 | 24.37 | 38.59 | 39.00 | 0.40*** | 24.0 |

| East Germany | (1) | (2) | (3) | (4) | | |
|---|---|---|---|---|---|---|
| Industry | N | CBA coverage (%) | Hours with CBA | Hours without CBA | Difference (4) - (3) | Diff. in minutes |
| All sectors | 63,010 | 13.28 | 39.24 | 39.58 | 0.34*** | 20.4 |
| Manufacturing | 22,083 | 11.50 | 39.08 | 39.71 | 0.63*** | 37.8 |
| Agriculture, mining, utilities and waste | 2,245 | 8.04 | 39.71 | 39.87 | 0.16*** | 9.6 |
| Construction | 5,163 | 38.52 | 39.67 | 39.78 | 0.11*** | 6.6 |
| Trade | 8,780 | 10.43 | 39.29 | 39.65 | 0.36*** | 21.6 |
| Transport, logistics, information and communication | 3,536 | 5.33 | 39.27 | 39.74 | 0.46*** | 27.6 |
| Hospitality, food and beverages | 5,542 | 9.74 | 39.48 | 39.57 | 0.09 | 5.4 |
| Finance, insurance and business services | 9,394 | 12.14 | 38.58 | 39.55 | 0.97*** | 58.2 |
| Education, health and other services | 6,267 | 8.41 | 39.05 | 39.24 | 0.19*** | 11.4 |

Notes: The table reports collective bargaining agreement (CBA) coverage and average agreed weekly working hours for full-time employees by industry and region. CBA coverage indicates whether the establishment is covered by a collective bargaining agreement. N denotes unweighted establishment-year observations used to calculate the weighted averages; all percentages, means, and differences are weighted. The difference is calculated by subtracting weekly hours in establishments with CBA coverage from weekly hours in establishments without CBA coverage and is converted into minutes by multiplying by 60. Statistical significance is based on t-tests of the difference between hours with and without CBA coverage using Stata's lincom command. * $p < .10$, ** $p < .05$, *** $p < .01$. Source: IAB-Establishment Panel, waves 2008-2024.

We then move beyond cross-sectional differences between covered and uncovered establishments and examine working-time changes around exits from collective bargaining coverage by comparing average weekly working hours in the years before and after exit. Tables OA2 and OA3 show sector specific results.

Across individual industries, statistically significant changes after exiting collective bargaining coverage are observed in West Germany (Table OA2) only in trade and in finance, insurance and business services, while changes in hospitality, food and beverages are only weakly statistically significant. In East Germany (Table OA3), a statistically significant increase is found only in finance, insurance and business services. Overall, these findings suggest that working-time adjustments after exit from collective bargaining coverage are limited in magnitude, although the estimates for certain industries, such as finance, insurance and business services appear not only statistically significant but also economically relevant.

**Table OA2. Agreed weekly working hours before and after exiting CBA coverage by industry, West Germany**

| West Germany | (1) | (2) | (3) | (4) | | |
|---|---|---|---|---|---|---|
| Industry | Always CBA | Never CBA | Before exit | After exit | Difference (4) - (3) | Diff. in minutes |
| All sectors | 38.96 | 39.38 | 39.07 | 39.19 | 0.12*** | 7.2 |
| N | 28,303 | 53,002 | 3,079 | 4,209 | | |
| N-Establishments | 8,796 | 18,141 | 1,265 | 1,228 | | |
| Manufacturing | 38.34 | 39.27 | 38.73 | 38.86 | 0.12 | 7.2 |
| N | 8,152 | 11,219 | 582 | 816 | | |
| N-Establishments | 2,309 | 3,628 | 229 | 219 | | |
| Agriculture, mining, utilities and waste | 39.39 | 39.66 | 39.84 | 39.90 | 0.06 | 3.6 |
| N | 1,177 | 1,375 | 130 | 166 | | |
| N-Establishments | 373 | 527 | 55 | 51 | | |
| Construction | 39.44 | 39.50 | 39.58 | 39.44 | -0.14 | -8.4 |
| N | 4,426 | 2,249 | 431 | 391 | | |
| N-Establishments | 1,237 | 954 | 148 | 131 | | |
| Trade | 38.88 | 39.38 | 38.94 | 39.15 | 0.20** | 12.0 |
| N | 4,349 | 10,251 | 762 | 1,141 | | |
| N-Establishments | 1,345 | 3,484 | 301 | 295 | | |
| Transport, logistics, information and communication | 39.00 | 39.55 | 39.48 | 39.38 | -0.10 | -6.0 |
| N | 1,270 | 4,982 | 160 | 237 | | |
| N-Establishments | 464 | 1,845 | 73 | 78 | | |
| Hospitality, food and beverages | 39.28 | 39.51 | 39.28 | 39.49 | 0.21* | 12.6 |
| N | 2,475 | 3,641 | 322 | 405 | | |
| N-Establishments | 796 | 1,419 | 137 | 133 | | |
| Finance, insurance and business services | 38.54 | 39.46 | 38.96 | 39.33 | 0.37*** | 22.2 |
| N | 4,276 | 12,092 | 319 | 483 | | |
| N-Establishments | 1,624 | 4,186 | 162 | 157 | | |
| Education, health and other services | 38.63 | 39.03 | 38.40 | 38.61 | 0.21 | 12.6 |
| N | 2,178 | 7,193 | 373 | 570 | | |
| N-Establishments | 915 | 2,649 | 179 | 183 | | |

Notes: The table reports average agreed weekly working hours of full-time employees in West Germany by industry and CBA coverage status. “Always CBA” and “Never CBA” denote establishments continuously covered and never covered by a CBA, respectively; “Before exit” and “After exit” refer to observations before and after leaving CBA coverage. N denotes unweighted establishment-year observations and N-Establishments the number of unique establishments used to calculate the weighted averages; all means and differences are weighted. Differences are calculated as hours without CBA minus hours with CBA and converted into minutes by multiplying by 60. Statistical significance is based on t-tests using Stata’s lincom command. * $p < .10$, ** $p < .05$, *** $p < .01$. Source: IAB-Establishment Panel, waves 2008-2024.

**Table OA3. Agreed weekly working hours before and after exiting CBA coverage by industry, East Germany**

| East Germany | (1) | (2) | (3) | (4) | | |
|---|---|---|---|---|---|---|
| Industry | Always CBA | Never CBA | Before exit | After exit | Difference (4) - (3) | Diff. in minutes |
| All sectors | 39.19 | 39.58 | 39.47 | 39.63 | 0.15*** | 9.0 |
| N | 10,539 | 48,177 | 1,591 | 2,703 | | |
| N-Establishments | 3,218 | 13,806 | 695 | 676 | | |
| Manufacturing | 38.99 | 39.72 | 39.42 | 39.44 | 0.02 | 1.2 |
| N | 2,956 | 17,658 | 511 | 958 | | |
| N-Establishments | 825 | 4,483 | 200 | 197 | | |
| Agriculture, mining, utilities and waste | 39.72 | 39.87 | 39.67 | 39.98 | 0.31 | 18.6 |
| N | 371 | 1,705 | 46 | 123 | | |
| N-Establishments | 96 | 466 | 25 | 23 | | |
| Construction | 39.65 | 39.78 | 39.76 | 39.78 | 0.01 | 0.6 |
| N | 2,157 | 2,345 | 315 | 346 | | |
| N-Establishments | 606 | 780 | 131 | 115 | | |
| Trade | 39.19 | 39.64 | 39.76 | 39.81 | 0.04 | 2.4 |
| N | 1,013 | 7,141 | 176 | 450 | | |
| N-Establishments | 359 | 2,052 | 100 | 97 | | |
| Transport, logistics, information and communication | 39.11 | 39.73 | 39.94 | 39.99 | 0.05 | 3.0 |
| N | 352 | 3,019 | 47 | 118 | | |
| N-Establishments | 139 | 1,008 | 28 | 28 | | |
| Hospitality, food and beverages | 39.33 | 39.56 | 39.84 | 39.79 | -0.05 | -3.0 |
| N | 856 | 4,252 | 192 | 242 | | |
| N-Establishments | 265 | 1,443 | 72 | 73 | | |
| Finance, insurance and business services | 38.53 | 39.56 | 38.83 | 39.30 | 0.47** | 28.2 |
| N | 2,120 | 6,935 | 148 | 191 | | |
| N-Establishments | 700 | 2,285 | 65 | 63 | | |
| Education, health and other services | 39.06 | 39.24 | 39.02 | 39.20 | 0.18 | 10.8 |
| N | 714 | 5,122 | 156 | 275 | | |
| N-Establishments | 317 | 1,748 | 91 | 90 | | |

Notes: The table reports average agreed weekly working hours of full-time employees in East Germany by industry and CBA coverage status. “Always CBA” and “Never CBA” denote establishments continuously covered and never covered by a CBA, respectively; “Before exit” and “After exit” refer to observations before and after leaving CBA coverage. N denotes unweighted establishment-year observations and N-Establishments the number of unique establishments used to calculate the weighted averages; all means and differences are weighted. Differences are calculated as hours without CBA minus hours with CBA and converted into minutes by multiplying by 60. Statistical significance is based on t-tests using Stata’s lincom command. * p < .10, ** p < .05, *** p < .01. Source: IAB-Establishment Panel, waves 2008-2024.

Taken together, the results point to some sectoral heterogeneity in the relationship between collective bargaining coverage and agreed working time. Rather than indicating a uniform adjustment of working hours after establishments leave collective bargaining coverage, the evidence suggests that adjustment patterns vary to some extent across sectors. This may reflect differences in the strength and reach of sectoral bargaining institutions, as well as the role of generally binding collective agreements, which can extend collectively agreed standards to establishments not formally covered by such agreements.

***Sector analysis: Empirical results***

We begin by reporting pooled OLS estimates that document differences in contractual weekly working hours between establishments covered by collective bargaining and those without coverage in West and East Germany by sector. The purpose of this exercise is primarily descriptive: it allows us to relate our findings to the existing literature on working-time differences by bargaining status.

The industry-specific results in Figure OA1 shows, that the working-time gap between covered and uncovered establishments varies substantially across sectors, particularly in West Germany. In West Germany, uncovered establishments report around 3 per cent longer weekly working hours in manufacturing and in finance, insurance and business services. This corresponds to differences of roughly 70 and 80 minutes per week, respectively. By contrast, no statistically significant difference is observed in construction. One plausible explanation is that working-time standards in this sector are shaped by collectively agreed rules that are extended by law to non-covered firms, thereby reducing differences between establishments with and without formal bargaining coverage.[10] In East Germany, cross-industry variation is less pronounced, although the estimated working-time gaps are broadly similar in magnitude to those observed in West Germany. Overall, the pooled OLS results confirm the finding of our descriptive results that collective bargaining coverage is associated with shorter agreed weekly working hours. At the same time, the industry-level evidence shows that this relationship is far from uniform. The working-time gap varies considerably across sectors and is absent in some industries, suggesting that the association between collective bargaining and working time depends strongly on the institutional and sectoral context.

[10] The Federal Framework Collective Agreement for the Construction Industry (Bundesrahmentarifvertrag für das Baugewerbe, BRTV), which has been declared generally binding, stipulates a weekly working time. of 38 hours for January to March and December, and 41 hours for April to November (§ 3 BRTV). For the extension of collective agreements in construction, including sectoral wage floors, see Möller et al. (2011).

**Figure OA1. Industry-specific POLS estimates of the association between non collective bargaining agreement coverage and agreed weekly working time among full-time employees in West and East Germany**

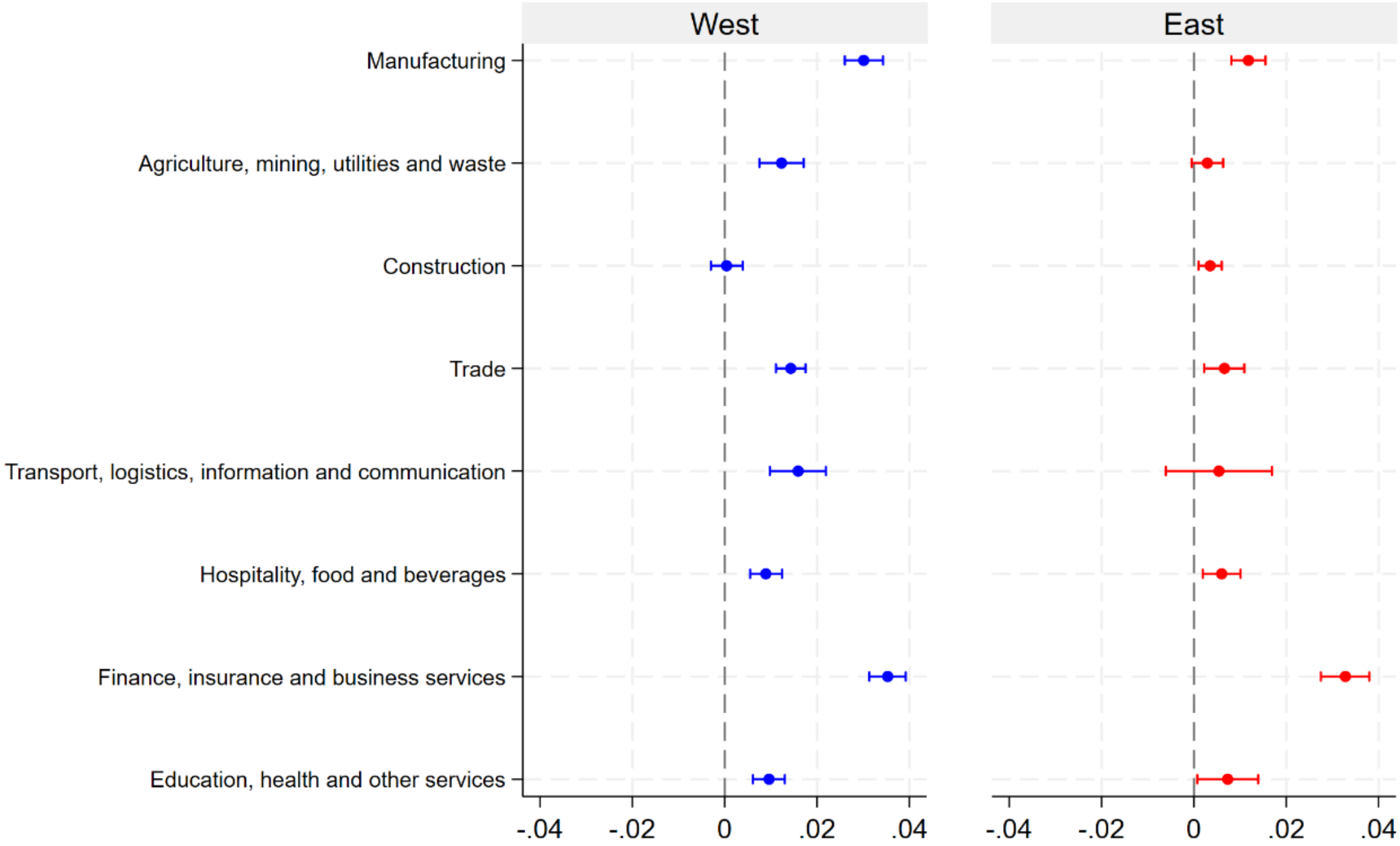

Notes: The figure shows industry-specific estimates of the association between non-collective bargaining agreement coverage and agreed weekly working time among full-time employees for West and East Germany. The dependent variable is the logarithm of agreed weekly working time for full-time employees. The figure reports coefficients and 95% confidence intervals. Standard errors are clustered at the establishment level. All models include year and federal-state fixed effects, establishment size controls, workforce composition controls, and establishment-level controls. Workforce composition controls include the shares of female, fixed-term, qualified, and part-time employees; establishment-level controls include lagged works council presence, previous-year profit situation, management structure, state of technical equipment, and an indicator for overtime in the previous year. Industry-specific sample sizes are: manufacturing (West: 13,705; East: 15,993); agriculture, mining, utilities and waste (West: 1,857; East: 1,658); construction (West: 5,009; East: 3,796); trade (West: 10,936; East: 6,399); transport, logistics, information and communication (West: 4,171; East: 2,388); hospitality, food and beverages (West: 4,285; East: 3,597); finance, insurance and business services (West: 10,957; East: 6,360); education, health and other services (West: 6,347; East: 3,989). Source: IAB-Establishment Panel, waves 2008-2024.

We next move beyond differences between covered and uncovered establishments and examine how exits from collective bargaining coverage are associated with changes in contractual working hours. For this purpose, we estimate fixed-effects models.

Figure OA2 shows that the fixed effect estimates indicate less pronounced sectoral heterogeneity than the pooled OLS results. In West Germany, most industry-specific estimates are statistically insignificant. The main exception is trade, where exit from collective bargaining coverage is associated with an increase in contractual weekly working hours of approximately 0.48 per cent, statistically significant at the five per cent level. Notably, the construction sector again displays no measurable effect, which is consistent with the institutional setting in this sector, where generally binding collective agreements may standardise working-time rules across establishments irrespective of formal bargaining coverage.

In East Germany, most industry-specific estimates are statistically insignificant and close to zero. The exceptions are manufacturing and finance, insurance and business services. In manufacturing, exit from collective bargaining coverage is associated with an increase in contractual weekly working hours of approximately 0.24 per cent. The largest estimate is found for finance, insurance and business services, where working hours increase by approximately 1.62 per cent. Evaluated at the mean weekly working time of covered establishments in this sector, 38.57 hours, this corresponds to an increase of around 38 minutes per week.

**Figure OA2. Industry-specific fixed-effects estimates of exits from collective bargaining coverage and agreed weekly working time among full-time employees in West and East Germany**

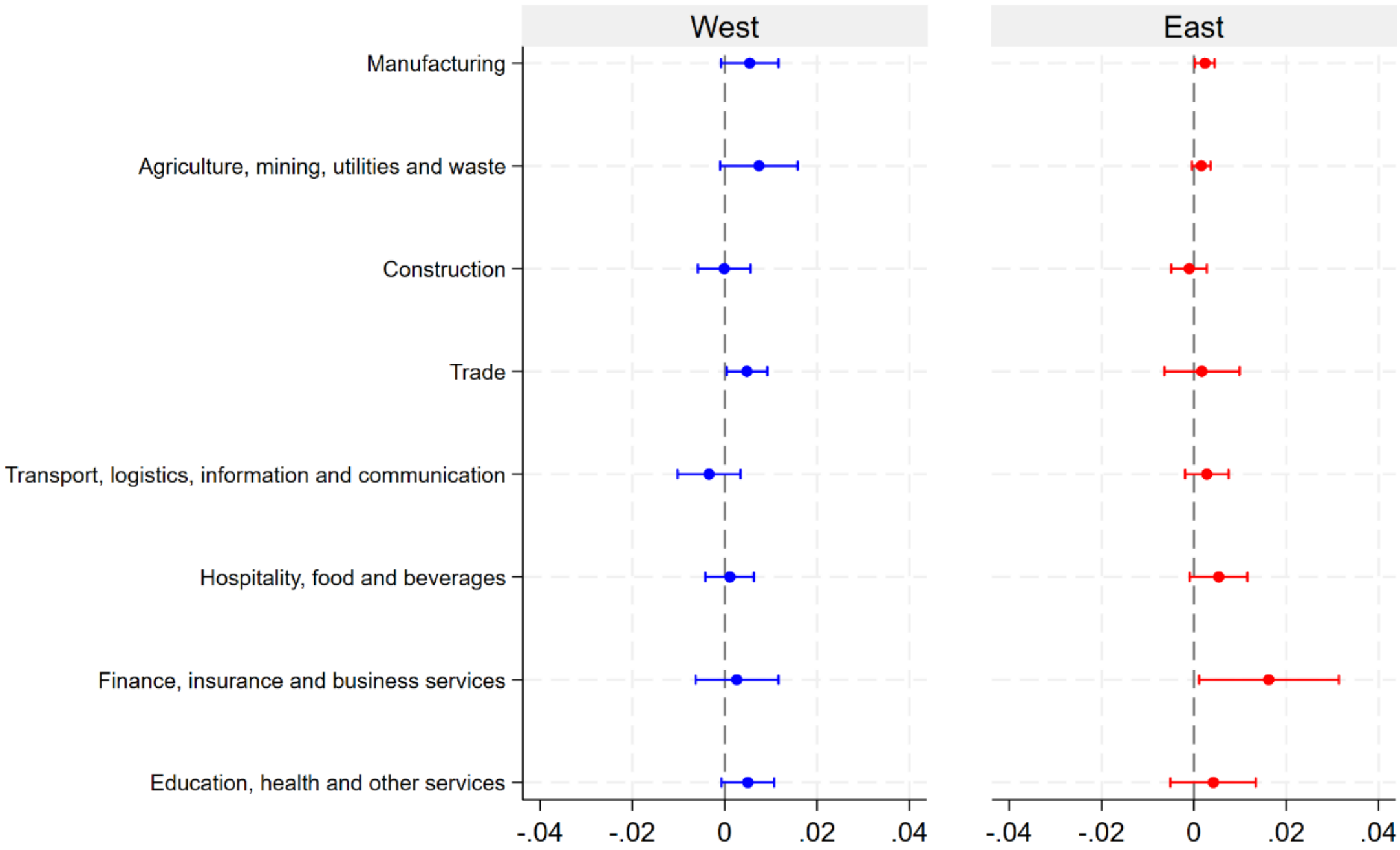

Notes: The figure shows industry-specific fixed-effects estimates of the association between non-collective bargaining agreement coverage and agreed weekly working time among full-time employees for West and East Germany. The dependent variable is the logarithm of agreed weekly working time for full-time employees. The figure reports coefficients and 95% confidence intervals. Standard errors are clustered at the establishment level. Establishments entering collective bargaining agreement coverage during the observation period are excluded; within-establishment variation in coverage therefore reflects exits from coverage. All models include year fixed effects, establishment fixed effects, establishment size controls, workforce composition controls, and establishment-level controls. Workforce composition controls include the shares of female, fixed-term, qualified, and part-time employees; establishment-level controls include lagged works council presence, previous-year profit situation, management structure, state of technical equipment, and an indicator for overtime in the previous year. Industry-specific sample sizes are: manufacturing (West: 13,705; East: 15,993); agriculture, mining, utilities and waste (West: 1,857; East: 1,658); construction (West: 5,009; East: 3,796); trade (West: 10,936; East: 6,399); transport, logistics, information and communication (West: 4,171; East: 2,388); hospitality, food and beverages (West: 4,285; East: 3,597); finance, insurance and business services (West: 10,957; East: 6,360); education, health and other services (West: 6,347; East: 3,989). Source: IAB-Establishment Panel, waves 2008-2024.

**Discussion of robustness checks**

Tables OA4 and OA5 report the corresponding estimates when the upper bound for agreed average weekly working hours of full-time employees is extended to 48 hours. Figures OA3 and OA4 additionally present the event-study results for both West and East Germany. Results from the POLS and fixed-effects estimations for both parts of the country are very similar to those in the main specification. With regard to the POLS estimates, the significance levels remain unchanged and the point estimates are almost identical. The same also holds for the fixed-effects estimates for West Germany. The only exception is the fixed-effects estimate for East Germany, where the coefficient is no longer statistically significant. However, with a coefficient of about 0.2 per cent, it still indicates a comparable association to that in the main specification, where the corresponding estimate is approximately 0.3 per cent. We therefore consider the POLS and fixed-effects results to be robust. The event-study estimates (Figures OA3 and OA4) show a similar pattern, although some coefficients lose statistical significance in both regions. However, as the confidence intervals overlap substantially and the point estimates are similar in magnitude to those of the main specification, we interpret the results as robust.

**Table OA4. Pooled OLS estimates of non-collective bargaining coverage and agreed weekly working time among full-time employees (35-48 hours) in West and East Germany**

| | West | East |
|---|---|---|
| Non-CBA coverage (yes = 1; no =0) | 0.0214***<br>(0.0009) | 0.0129***<br>(0.0011) |
| Year FE | yes | yes |
| Industry FE | yes | yes |
| Federal-state FE | yes | yes |
| Establishment FE | no | no |
| Adjusted $R^2$ | 0.222 | 0.082 |
| Observations | 60,649 | 46,077 |

Notes: The table reports pooled ordinary least squares estimates for West and East Germany. The dependent variable is the logarithm of agreed weekly working time for full-time employees (35-48 hours). Collective bargaining agreement (CBA) coverage indicates whether the establishment is not covered by a collective bargaining agreement. Standard errors, clustered at the establishment level, are reported in parentheses. All models include year fixed effects, federal-state fixed effects, industry fixed effects, establishment size controls, workforce composition controls, and additional establishment-level controls. Workforce composition controls include the shares of female, fixed-term, qualified, and part-time employees. Establishment-level controls include lagged works council presence, profit situation in the previous year, management structure, state of technical equipment, and an indicator for overtime in the previous year. * p < .10, ** p < .05, *** p < .01. Source: IAB-Establishment Panel, waves 2008-2024.

**Table OA5. Fixed-effects estimates of exits from collective bargaining coverage and agreed weekly working time among full-time employees (35-48 hours) in West and East Germany**

| | West | East |
|---|---|---|
| Non-CBA coverage (yes = 1; no =0) | 0.0035***<br>(0.0013) | 0.0019<br>(0.0017) |
| Year FE | yes | yes |
| Industry FE | yes | yes |
| Federal-state FE | yes | yes |
| Establishment FE | yes | yes |
| Adjusted $R^2$ | 0.002 | 0.011 |
| Observations | 60,649 | 46,077 |

Notes: The table reports fixed-effects estimates for West and East Germany. The dependent variable is the logarithm of agreed weekly working time for full-time employees (35-48 hours). Collective bargaining agreement (CBA) coverage indicates whether the establishment is not covered by a collective bargaining agreement. Establishments entering collective bargaining agreement coverage during the observation period are excluded; therefore, within-establishment variation in coverage reflects exits from coverage. As a result, the coefficient indicates an increase in agreed weekly working time after exiting collective bargaining agreement coverage. Standard errors, clustered at the establishment level, are reported in parentheses. All models include year fixed effects, establishment fixed effects, establishment size controls, workforce composition controls, and additional establishment-level controls. Workforce composition controls include the shares of female, fixed-term, qualified, and part-time employees. Establishment-level controls include lagged works council presence, profit situation in the previous year, management structure, state of technical equipment, and an indicator for overtime in the previous year. * p < .10, ** p < .05, *** p < .01. Source: IAB-Establishment Panel, waves 2008-2024.

**Figure OA3. Event-study estimates of agreed weekly working time among full-time employees (35-48 hours) around exit from collective bargaining coverage in West Germany**

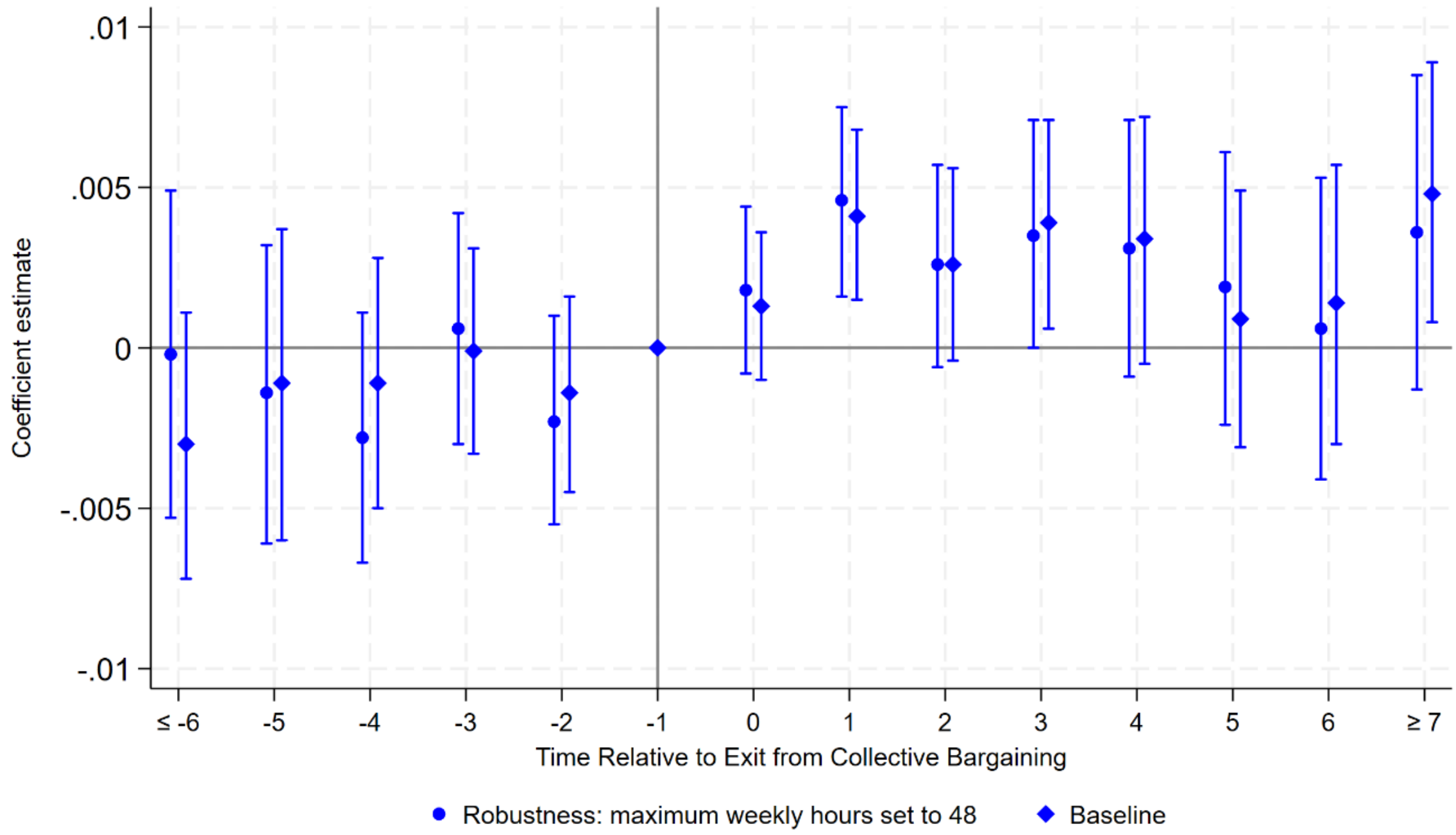

Notes: The figure reports event-study estimates of the change in agreed weekly working time around exit from collective bargaining coverage for West Germany. The dependent variable is the logarithm of agreed weekly working time for full-time employees (35-48 hours). Coefficients are measured relative to the year before exit from collective bargaining coverage. Points show coefficient estimates and vertical lines show 95% confidence intervals. The event-time categories ≤ -6 and ≥ 7 pool observations six or more years before and seven or more years after exit, respectively. The baseline specification includes establishments with agreed weekly working time of 35-40 hours among full-time employees. Standard errors are clustered at the establishment level. Models include establishment fixed effects, year fixed effects, federal-state fixed effects, establishment size controls, workforce composition controls, and establishment-level controls. Workforce composition controls include the shares of female, fixed-term, qualified, and part-time employees; establishment-level controls include lagged works council presence, previous-year profit situation, management structure, state of technical equipment, and an indicator for overtime in the previous year. The estimation sample of the robustness check includes 822,935 establishment-year observations and the baseline model 778,001. Source: IAB-Establishment Panel, waves 2008-2024.

**Figure OA4. Event-study estimates of agreed weekly working time among full-time employees (35-48 hours) around exit from collective bargaining coverage in East Germany**

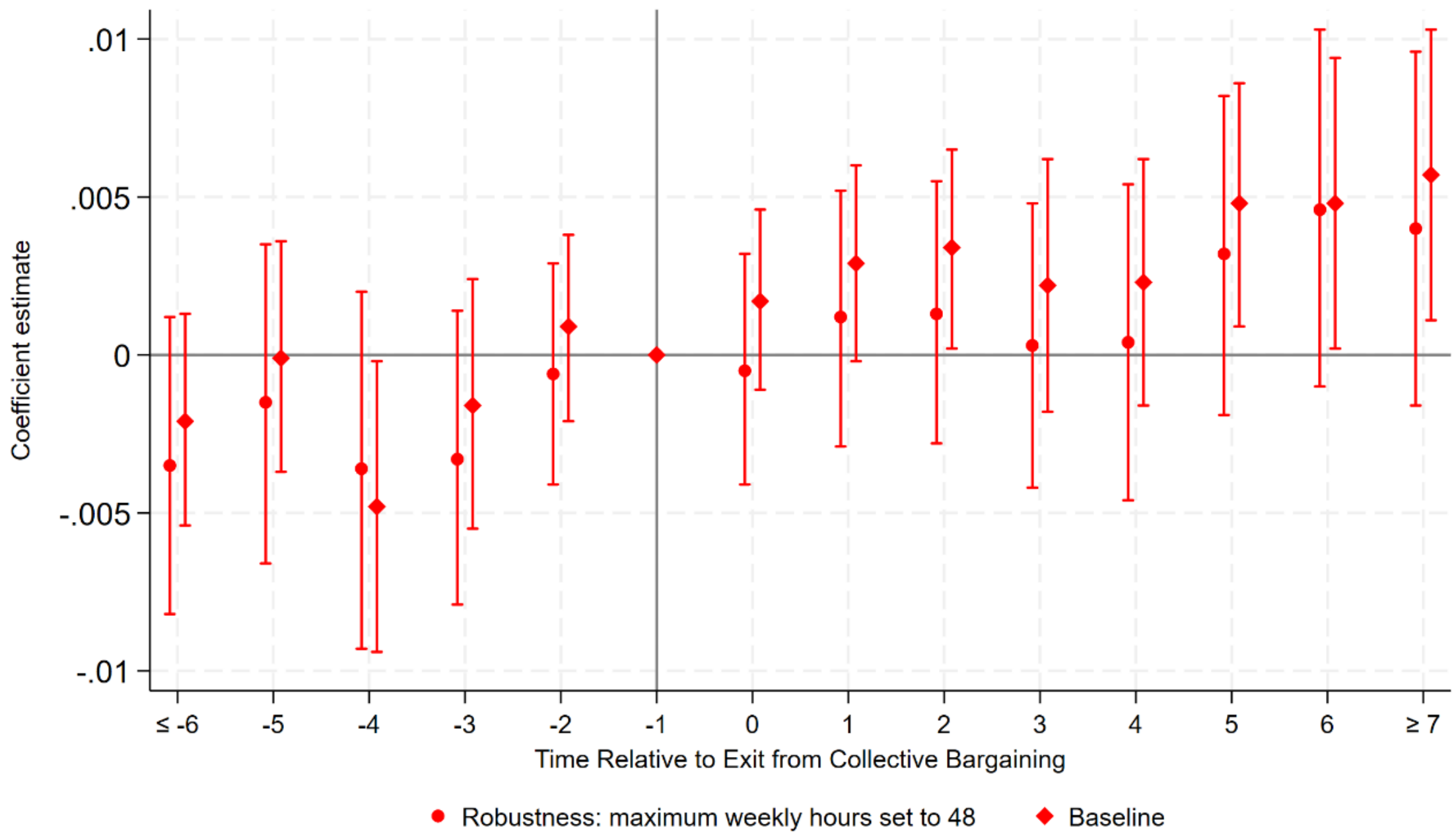

Notes: The figure reports event-study estimates of the change in agreed weekly working time around exit from collective bargaining coverage for East Germany. The dependent variable is the logarithm of agreed weekly working time for full-time employees (35-48 hours). Coefficients are measured relative to the year before exit from collective bargaining coverage. Points show coefficient estimates and vertical lines show 95% confidence intervals. The event-time categories ≤ -6 and ≥ 7 pool observations six or more years before and seven or more years after exit, respectively. The baseline specification includes establishments with agreed weekly working time of 35-40 hours among full-time employees. Standard errors are clustered at the establishment level. Models include establishment fixed effects, year fixed effects, federal-state fixed effects, establishment size controls, workforce composition controls, and establishment-level controls. Workforce composition controls include the shares of female, fixed-term, qualified, and part-time employees; establishment-level controls include lagged works council presence, previous-year profit situation, management structure, state of technical equipment, and an indicator for overtime in the previous year. The estimation sample of the robustness check includes 639,565 establishment-year observations and the baseline model 614,330. Source: IAB-Establishment Panel, waves 2008-2024.

Excluding the years from 2020 onwards does not materially affect the POLS or fixed-effects estimates for West and East Germany. The POLS estimates (Table OA6) are nearly identical in both parts of Germany with respect to both the size of the coefficients and their statistical significance. The fixed-effects estimates (Table OA7) for West and East Germany remain statistically significant, although now at the five per cent level rather than the one per cent level. For both parts of Germany, the point estimates are somewhat smaller but still comparable to those in the main specification. The event-study estimates (Figures OA5 and OA6) show a broadly similar overall pattern to the main specification, although some coefficients lose statistical significance. We therefore conclude that our findings are not driven by the Covid-19 pandemic.

**Table OA6. Pooled OLS estimates of non-collective bargaining coverage and agreed weekly working time among full-time employees in West and East Germany, 2008-2019**

| | West | East |
|---|---|---|
| Non-CBA coverage (yes = 1; no =0) | 0.0210***<br>(0.0009) | 0.0135***<br>(0.0011) |
| Year FE | yes | yes |
| Industry FE | yes | yes |
| Federal-state FE | yes | yes |
| Establishment FE | no | no |
| Adjusted $R^2$ | 0.273 | 0.103 |
| Observations | 40,997 | 31,158 |

Notes: The table reports pooled ordinary least squares estimates for West and East Germany, covering the period 2008-2019. The dependent variable is the logarithm of agreed weekly working time for full-time employees. Collective bargaining agreement (CBA) coverage indicates whether the establishment is not covered by a collective bargaining agreement. Standard errors, clustered at the establishment level, are reported in parentheses. All models include year fixed effects, federal-state fixed effects, industry fixed effects, establishment size controls, workforce composition controls, and additional establishment-level controls. Workforce composition controls include the shares of female, fixed-term, qualified, and part-time employees. Establishment-level controls include lagged works council presence, profit situation in the previous year, management structure, state of technical equipment, and an indicator for overtime in the previous year. * p < .10, ** p < .05, *** p < .01. Source: IAB-Establishment Panel, waves 2008-2019.

**Table OA7. Fixed-effects estimates of exits from collective bargaining coverage and agreed weekly working time among full-time employees in West and East Germany, 2008-2019**

| | West | East |
|---|---|---|
| Non-CBA coverage (yes = 1; no =0) | 0.0027** (0.0013) | 0.0025** (0.0010) |
| Year FE | yes | yes |
| Industry FE | yes | yes |
| Federal-state FE | yes | yes |
| Establishment FE | yes | yes |
| Adjusted $R^2$ | 0.002 | 0.007 |
| Observations | 40,997 | 31,158 |

Notes: The table reports fixed-effects estimates for West and East Germany, covering the period 2008-2019. The dependent variable is the logarithm of agreed weekly working time for full-time employees. Collective bargaining agreement (CBA) coverage indicates whether the establishment is not covered by a collective bargaining agreement. Establishments entering collective bargaining agreement coverage during the observation period are excluded; therefore, within-establishment variation in coverage reflects exits from coverage. As a result, the coefficient indicates an increase in agreed weekly working time after exiting collective bargaining agreement coverage. Standard errors, clustered at the establishment level, are reported in parentheses. All models include year fixed effects, establishment fixed effects, establishment size controls, workforce composition controls, and additional establishment-level controls. Workforce composition controls include the shares of female, fixed-term, qualified, and part-time employees. Establishment-level controls include lagged works council presence, profit situation in the previous year, management structure, state of technical equipment, and an indicator for overtime in the previous year. * p < .10, ** p < .05, *** p < .01. Source: IAB-Establishment Panel, waves 2008-2019.

**Figure OA5. Event-study estimates of agreed weekly working time among full-time employees around exit from collective bargaining coverage in West Germany, 2008-2019**

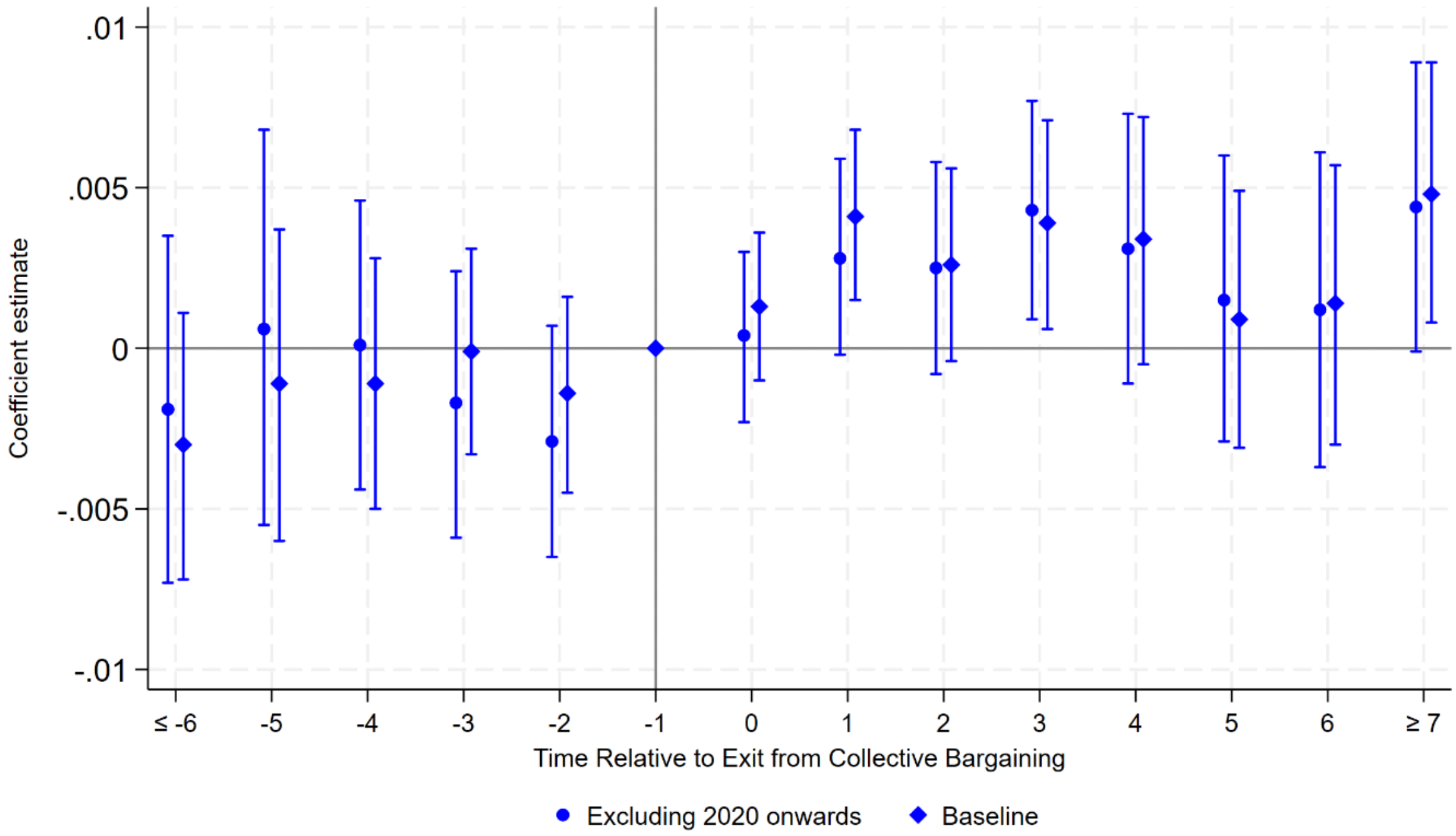

Notes: The figure reports event-study estimates of the change in agreed weekly working time around exit from collective bargaining coverage for West Germany, covering the period 2008-2019. The dependent variable is the logarithm of agreed weekly working time for full-time employees. Coefficients are measured relative to the year before exit from collective bargaining coverage. Points show coefficient estimates and vertical lines show 95% confidence intervals. The event-time categories ≤ -6 and ≥ 7 pool observations six or more years before and seven or more years after exit, respectively. The baseline specification includes establishments with agreed weekly working time of 35-40 hours among full-time employees. Standard errors are clustered at the establishment level. Models include establishment fixed effects, year fixed effects, federal-state fixed effects, establishment size controls, workforce composition controls, and establishment-level controls. Workforce composition controls include the shares of female, fixed-term, qualified, and part-time employees; establishment-level controls include lagged works council presence, previous-year profit situation, management structure, state of technical equipment, and an indicator for overtime in the previous year. The estimation sample of the robustness check includes 368,634 establishment-year observations and the baseline model 778,001. Source: IAB-Establishment Panel, waves 2008-2019.

**Figure OA6. Event-study estimates of agreed weekly working time among full-time employees around exit from collective bargaining coverage in East Germany, 2008-2019**

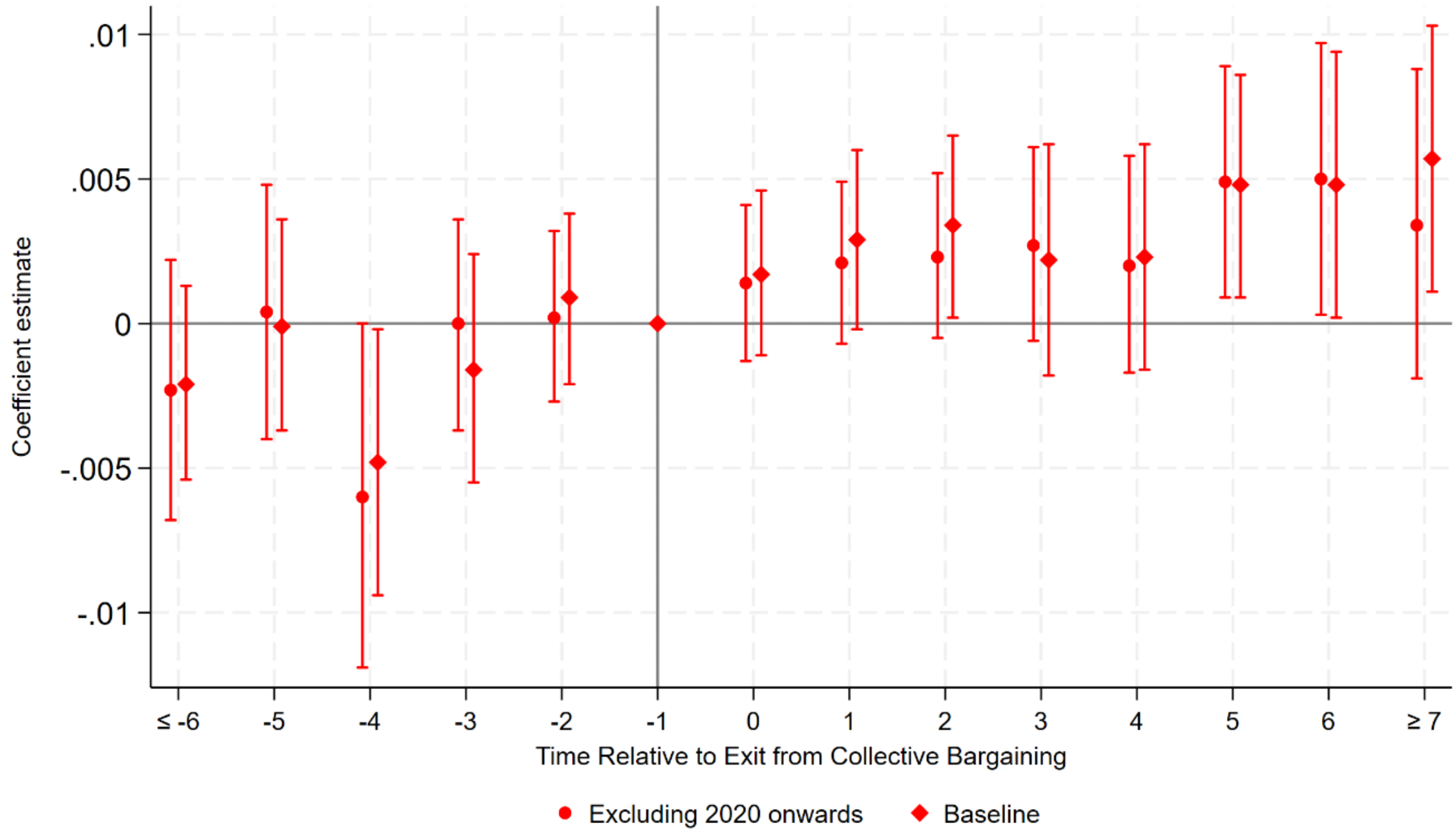

Notes: The figure reports event-study estimates of the change in agreed weekly working time around exit from collective bargaining coverage for East Germany, covering the period 2008-2019. The dependent variable is the logarithm of agreed weekly working time for full-time employees. Coefficients are measured relative to the year before exit from collective bargaining coverage. Points show coefficient estimates and vertical lines show 95% confidence intervals. The event-time categories ≤ -6 and ≥ 7 pool observations six or more years before and seven or more years after exit, respectively. The baseline specification includes establishments with agreed weekly working time of 35-40 hours among full-time employees. Standard errors are clustered at the establishment level. Models include establishment fixed effects, year fixed effects, federal-state fixed effects, establishment size controls, workforce composition controls, and establishment-level controls. Workforce composition controls include the shares of female, fixed-term, qualified, and part-time employees; establishment-level controls include lagged works council presence, previous-year profit situation, management structure, state of technical equipment, and an indicator for overtime in the previous year. The estimation sample of the robustness check includes 287,343 establishment-year observations and the baseline model 614,330. Source: IAB-Establishment Panel, waves 2008-2019.

When considering only federal-state, industry, and year fixed effects in our POLS estimates (Table OA8), the results remain similar in terms of both statistical significance and the magnitude of the point estimates, although the differences in the coefficients are slightly larger in West Germany. The results from the fixed-effects estimations (see also Table OA8) as well as the event-study results (Figures OA7 and OA8) are also consistent across both specifications. We therefore conclude that bad controls do not appear to pose a problem in our estimations.

**Table OA8. Pooled OLS estimates of non-collective bargaining agreement coverage and agreed weekly working time among full-time employees in West and East Germany**

| | West | | East | |
|---|---|---|---|---|
| | POLS | FE | POLS | FE |
| Non-CBA coverage (yes = 1; no =0) | 0.0277***<br>(0.0006) | 0.0029***<br>(0.0008) | 0.0152***<br>(0.0007) | 0.0027***<br>(0.0008) |
| Year FE | yes | yes | yes | yes |
| Industry FE | yes | yes | yes | yes |
| Federal-state FE | yes | yes | yes | yes |
| Establishment FE | no | yes | no | yes |
| Adjusted $R^2$ | 0.173 | 0.002 | 0.072 | 0.008 |
| Mean, CBA | 38.36 | 38.36 | 39.13 | 39.13 |
| Mean, non-CBA | 39.37 | 39.37 | 39.65 | 39.65 |
| Observations | 91,972 | 91,972 | 66,342 | 66,342 |
| CBA observations | 31,561 | 31,561 | 12,237 | 12,237 |
| Non-CBA observations | 60,411 | 60,411 | 54,105 | 54,105 |

Notes: The table reports pooled ordinary least squares as well as the fixed effects estimates for West and East Germany. The dependent variable is the logarithm of agreed weekly working time for full-time employees. Collective bargaining agreement (CBA) coverage indicates whether the establishment is not covered by a collective bargaining agreement. Standard errors, clustered at the establishment level, are reported in parentheses. All models include year fixed effects, federal-state fixed effects, industry fixed effects. * p < .10, ** p < .05, *** p < .01. Source: IAB-Establishment Panel, waves 2008-2024.

**Figure OA7. Event-study estimates of agreed weekly working time among full-time employees with reduced number of control variables in West Germany**

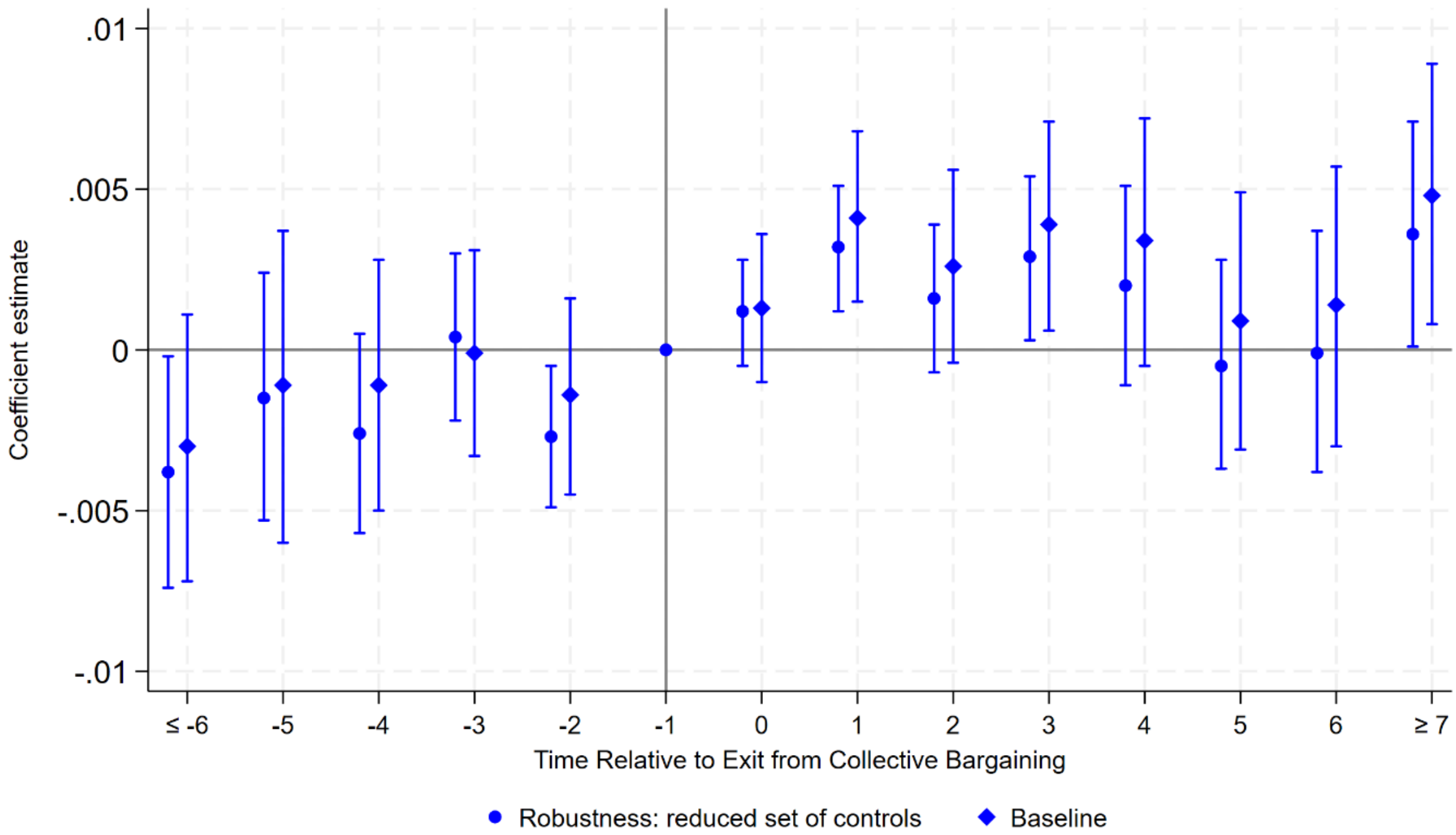

Notes: The figure reports event-study estimates of the change in agreed weekly working time around exit from collective bargaining coverage for West Germany. The dependent variable is the logarithm of agreed weekly working time for full-time employees. Coefficients are measured relative to the year before exit from collective bargaining coverage. Points show coefficient estimates and vertical lines show 95% confidence intervals. The event-time categories ≤ -6 and ≥ 7 pool observations six or more years before and seven or more years after exit, respectively. The baseline specification includes the full set of control variables: establishment fixed effects, year fixed effects, federal-state fixed effects, establishment size controls, workforce composition controls, and establishment-level controls. Workforce composition controls include the shares of female, fixed-term, qualified, and part-time employees; establishment-level controls include lagged works council presence, previous-year profit situation, management structure, state of technical equipment, and an indicator for overtime in the previous year. While the robustness check includes only establishment fixed effects, year fixed effects, federal-state fixed effects and sector fixed effect. The estimation sample of the robustness check includes 1,271,948 establishment-year observations and the baseline model 778,001. Source: IAB-Establishment Panel, waves 2008-2024.

**Figure OA8. Event-study estimates of agreed weekly working time among full-time employees with reduced number of control variables in East Germany**

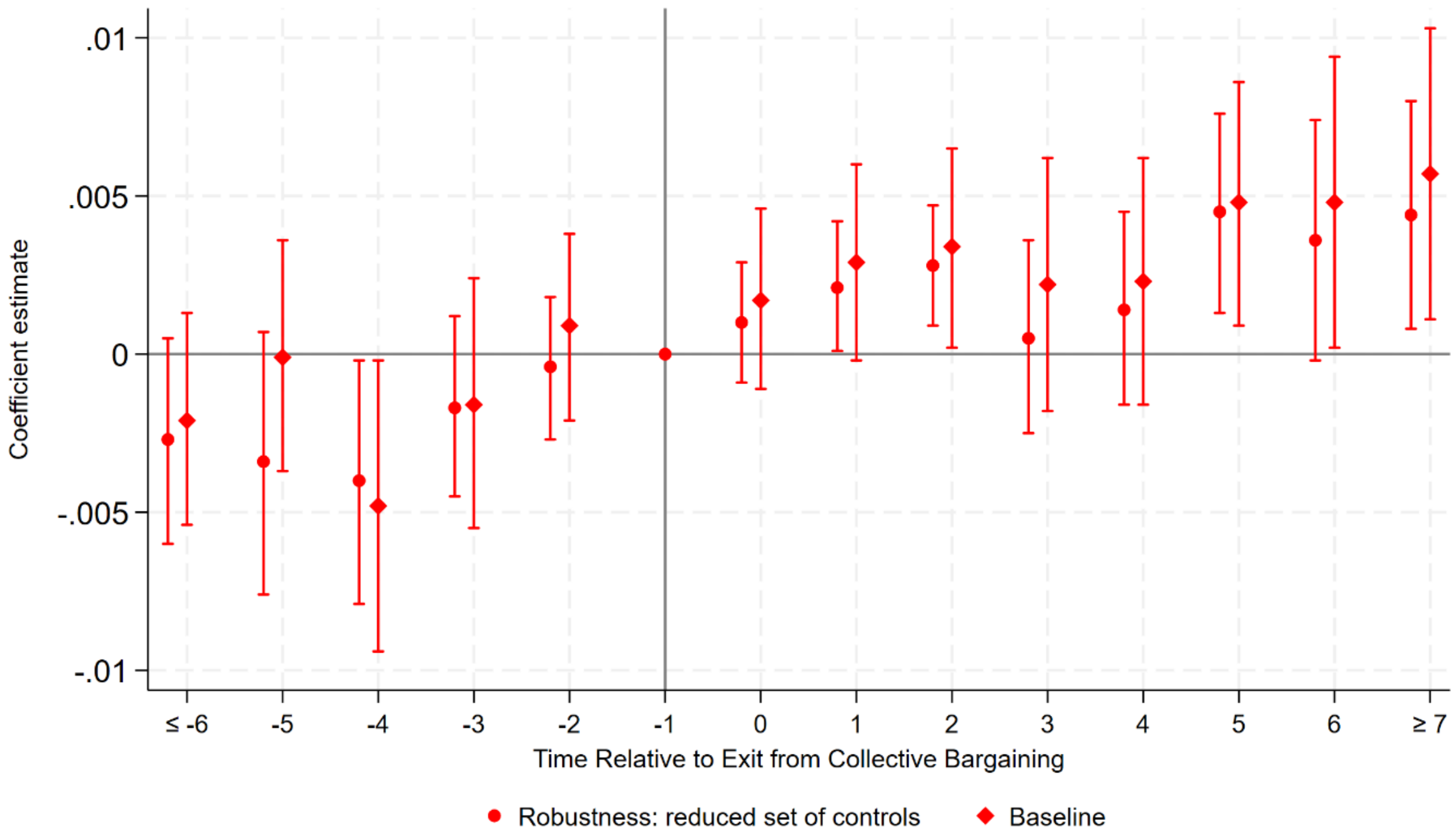

Notes: The figure reports event-study estimates of the change in agreed weekly working time around exit from collective bargaining coverage for East Germany. The dependent variable is the logarithm of agreed weekly working time for full-time employees. Coefficients are measured relative to the year before exit from collective bargaining coverage. Points show coefficient estimates and vertical lines show 95% confidence intervals. The event-time categories ≤ -6 and ≥ 7 pool observations six or more years before and seven or more years after exit, respectively. The baseline specification includes the full set of control variables: establishment fixed effects, year fixed effects, federal-state fixed effects, establishment size controls, workforce composition controls, and establishment-level controls. Workforce composition controls include the shares of female, fixed-term, qualified, and part-time employees; establishment-level controls include lagged works council presence, previous-year profit situation, management structure, state of technical equipment, and an indicator for overtime in the previous year. While the robustness check includes only establishment fixed effects, year fixed effects, federal-state fixed effects and sector fixed effect. The estimation sample of the robustness check includes 932,452 establishment-year observations and the baseline model 614,330. Source: IAB-Establishment Panel, waves 2008-2024.

**Figure OA9. Callaway and Sant´Anna (2021) estimator of agreed weekly working time among full-time employees.**

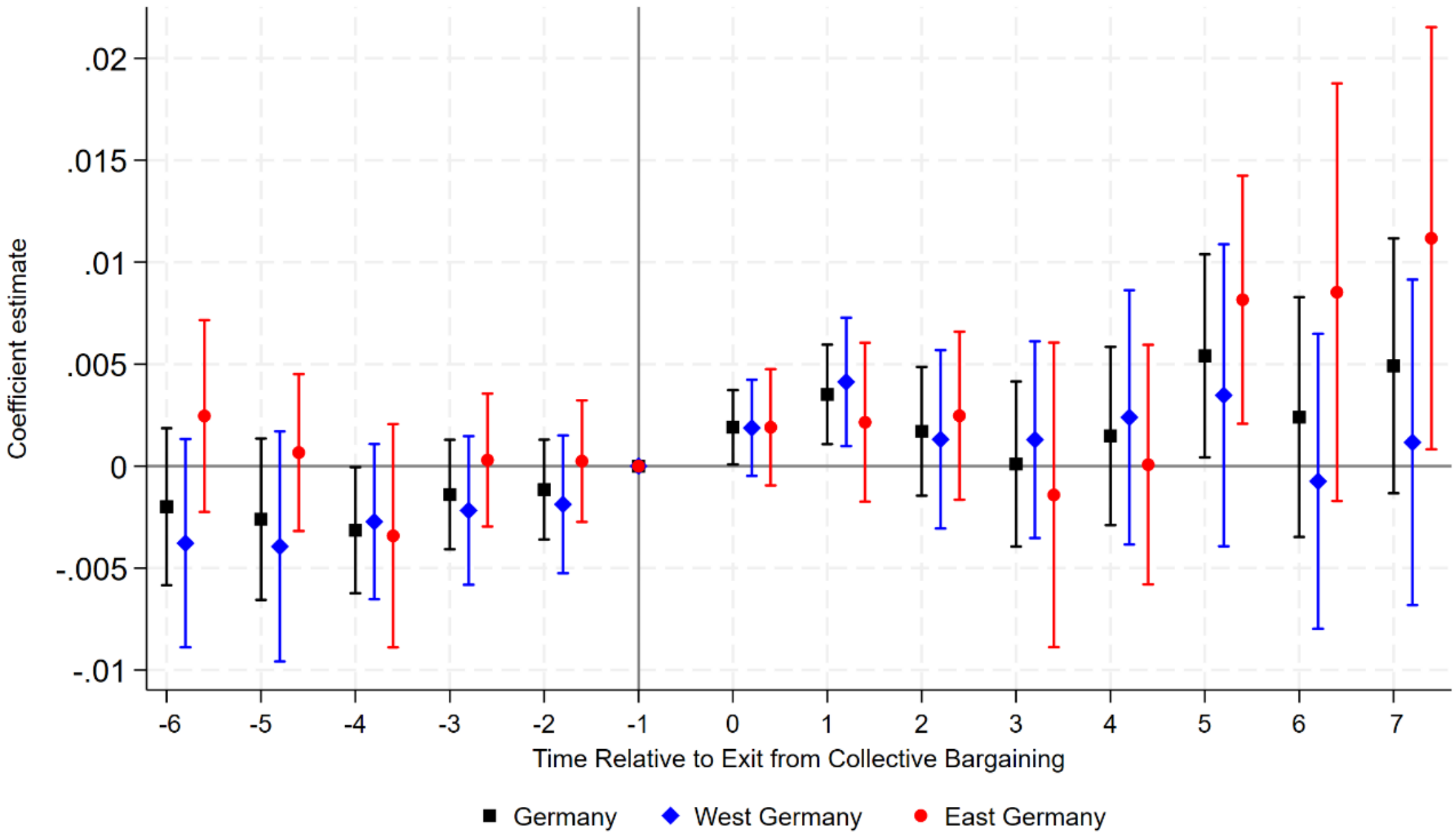

Notes: The figure reports event-study estimates using Callaway and Sant´Anna (2021) of the change in agreed weekly working time around exit from collective bargaining coverage for Germany, West Germany and East Germany. The dependent variable is the logarithm of agreed weekly working time for full-time employees. Coefficients are measured relative to the year before exit from collective bargaining coverage. Points show coefficient estimates and vertical lines show 95% confidence intervals. Standard errors are clustered at the establishment level. Models include establishment fixed effects, year fixed effects, federal-state fixed effects, establishment size controls, workforce composition controls, and establishment-level controls. Workforce composition controls include the shares of female, fixed-term, qualified, and part-time employees; establishment-level controls include lagged works council presence, previous-year profit situation, management structure, state of technical equipment, and an indicator for overtime in the previous year. The estimation is done by using the regression DiD estimator based on ordinary least squares. Sample includes 88,499 establishment-year observations in Germany, 49,485 in West Germany and 39,011 in East Germany. Source: IAB-Establishment Panel, waves 2008-2024.

**In der Diskussionspapierreihe sind kürzlich erschienen:**

**Recently published Discussion Papers:**

| | | | |
|---|---|---|---|
| 135 | Pyka, V., Rieder, A. | The Myth of Restrictive Working-Time Agreements. Firms' Working-Time Adjustment after Exit from Collective Bargaining | 09/2026 |
| 134 | Weikl, J. | Preference-Driven Contract Design: How Education Alters Risk, Patience, and Effort in Incentive Schemes | 01/2026 |
| 133 | Rieder, A. Schnabel, C. | Beyond Collective Agreements: The Rise of the Wage Cushion in Germany | 09/2025 |
| 132 | Pyka, V. | Pay Clauses in Public Procurement: The Wage Impact of Collective Bargaining Compliance Laws in Germany | 06/2025 |
| 131 | Schnabel, C. | Unions and Employers' Associations in Germany: A Survey of their Membership, Density and Bargaining Coverage | 01/2025 |
| 130 | Kohaut, S., Schnabel, C. | The Demise of Works Councils in Germany | 05/2024 |
| 129 | Kovalenko, T., Schröpf, B. | Patterns of regional firm mobility in Germany | 04/2024 |
| 128 | Pyka, V., Schnabel, C. | Unionization of retired workers in Europe | 11/2023 |
| 127 | Wieser, L., Abraham, M., Schnabel, C., Niessen, C., Wolff, M. | When are Employers interested in Electronic Performance Monitoring? Results from a Factorial Survey Experiment | 06/2023 |
| 126 | Kovalenko, T., Sauerbier, T., Schröpf, B. | The fall and rebound of average establishment size in West Germany | 06/2023 |
| 125 | Bonaccolto-Töpfer, M., Schnabel, C. | Is There a Union Wage Premium in Germany and Which Workers Benefit Most? | 01/2023 |
| 124 | Bucher, F., Scheu, L., Schröpf, B. | Economic complexity and environmental pollution: Evidence from the former socialist transition countries | 03/2022 |

Eine aktualisierte Liste der Diskussionspapiere findet sich auf der Homepage:
http://www.arbeitsmarkt.rw.fau.de/

An updated list of discussion papers can be found at the homepage:
http://www.arbeitsmarkt.rw.fau.de